\documentclass[aps, prd, 10pt, twocolumn, superscriptaddress]{revtex4-1}

\usepackage{graphicx}
\usepackage{amssymb, amsmath,mathtools}
\usepackage{bbm}
\usepackage{enumerate}

\graphicspath{{./}{./plots/}}

\DeclareMathOperator{\Tr}{Tr}

\newcommand{\Nc}{N_{\mathrm{c}}}
\newcommand{\ii}{\mathrm{i}}
\newcommand{\ee}{\mathrm{e}}

\begin{document}

\title{Phase diagram of electronic systems with quadratic Fermi nodes in $2<d<4$:\\ $2+\epsilon$ expansion, $4-\epsilon$ expansion, and functional renormalization group}

\author{Lukas Janssen}
\affiliation{Institut f\"ur Theoretische Physik, Technische Universit\"at Dresden, 01062 Dresden, Germany}

\author{Igor F.\ Herbut}
\affiliation{Department of Physics, Simon Fraser University, Burnaby, British Columbia, Canada V5A 1S6}

\begin{abstract}

Several materials in the regime of strong spin-orbit interaction such as HgTe, the pyrochlore iridate Pr$_2$Ir$_2$O$_7$, and the half-Heusler compound LaPtBi, as well as various systems related to these three prototype materials, are believed to host a quadratic band touching point at the Fermi level.
Recently, it has been proposed that such a three-dimensional gapless state
is unstable to a Mott-insulating ground state at low temperatures when the number of band touching points $N$ at the Fermi level is smaller than a certain critical number $\Nc$.
We further substantiate and quantify this scenario by various approaches.
Using $\epsilon$ expansion near two spatial dimensions, we show that $\Nc = 64/(25 \epsilon^2) + \mathcal O(1/\epsilon)$ and demonstrate that the instability for $N<\Nc$ is towards a nematic ground state that can be understood as if the system were under (dynamically generated) uniaxial strain.
We also propose a truncation of the functional renormalization group equations in the dynamical bosonization scheme which we show to agree to one-loop order with the results from $\epsilon$ expansion both near two as well as near four dimensions, and which smoothly interpolates between these two perturbatively accessible limits for general $2<d<4$. Directly in $d=3$ we therewith find $\Nc = 1.86$, and thus again above the physical $N=1$.
All these results are consistent with the prediction that the interacting ground state of pure, unstrained HgTe, and possibly also Pr$_2$Ir$_2$O$_7$, is a strong topological insulator with a dynamically-generated gap---a topological Mott insulator.

\end{abstract}

\maketitle

\section{Introduction}

Solid matter is commonly classified by electrical transport behavior. In a sufficiently pure metal (or semimetal) the conductivity diverges when temperature $T$ goes to zero, or witnesses a superconducting transition at finite (though usually small)~$T$. On the other hand, we call a material an insulator (or semiconductor), if its conductivity vanishes when the system is cooled down. Scanning tunneling spectroscopy measurements correlate with this behavior: In a metal the density of states near the Fermi level is finite, while there is a finite gap in the spectrum of an insulator (or semiconductor).

Lately, a new class of materials that do not quite fit into this scheme has moved into the focus of attention: Systems in which the Fermi surface shrinks to a few isolated points in the Brillouin zone live right on the edge between metals and semiconductors. In the purest graphene samples, for instance, the conductivity at charge neutrality decreases with decreasing temperature (as in a semiconductor), but approaches at the lowest temperatures a finite value on the order of the conductance quantum $e^2 / h$~\cite{geim2007}. In three dimensions, materials with strong spin-orbit coupling can host pairs of Weyl nodes, which may be thought of as three-dimensional (3D) analogues of graphene's linear band crossing points, albeit with nontrivial topology~\cite{vafek2014}. Although in 3D the density of states as function of energy $\varepsilon$ vanishes as $D(\varepsilon) \sim \varepsilon^2$ near the linear band crossing point, the low-temperature conductivity is large as long as disorder is weak---an unusual (semi)metallic state~\cite{burkov2011}. Such a \textit{Weyl semimetal} state has recently been identified experimentally, together with its characteristic Fermi-arc surface states, in TaAs~\cite{xu2015, lv2015}.

In this work we focus on three-dimensional systems with \textit{quadratic} Fermi nodes, as realized, for instance, in $\alpha$-Sn or HgTe. These materials feature band inversion due to strong spin-orbit coupling, ensuring that the degeneracy at the quadratic band touching point (QBT) is protected by the crystal's cubic symmetry, and in the undoped situation the Fermi energy is right at the touching point~\cite{tsidilkovski1997}. Furthermore, no other states turn out to cross the Fermi level away from this point in these materials.
The systems may thus be viewed as 3D analogues of \textit{bilayer} graphene, however, with just a single QBT that is located at the center $\Gamma$ of the Brillouin zone. Such low-energy band structure has also been measured in the pyrochlore iridate Pr$_2$Ir$_2$O$_7$, therewith making it a strongly-correlated analogue of HgTe~\cite{kondo2015}. Recently, a quadratic Fermi node in 3D has also been identified in the related compound Nd$_2$Ir$_2$O$_7$~\cite{nakayama2016}.
In these materials, it is believed that the main physics is predominantly driven by the iridium electronic structure, while the interaction with the local rare-earth moments playing a role only at very low temperatures~\cite{witczakkrempa2014}.
Another family of materials which may host a QBT at the Fermi level are given by ternary half-Heusler compounds, with LaPtBi as a prototype system~\cite{chadov2010, lin2010, xiao2010}.

In the noninteracting case, the conductivity in a Fermi system with QBT vanishes at low temperatures with a power law~\cite{moon2013} (in contrast to the Weyl semimetal), and the system hence should properly be regarded as ``gapless semiconductor''~\cite{tsidilkovski1997}. However, as the density of states has the square-root form as function of energy $\varepsilon$, $D(\varepsilon) \sim \sqrt{\varepsilon}$, the long-range Coulomb interaction is only marginally screened and the low-temperature behavior of these systems might decisively depend on the role played by the interactions. In fact, using an $\epsilon$ expansion around the upper critical spatial dimension of $d=4$, as well as in the limit of large number $N$ of QBTs at the Fermi level, Abrikosov and Beneslavskii~\cite{abrikosov1971, abrikosov1974} found a scale-invariant interacting ground state with unusual power laws in various thermodynamic observables---a 3D non-Fermi liquid (NFL) state. This scenario has recently been reviewed and put forward as an explanation for the anomalous low-temperature behavior measured in Pr$_2$Ir$_2$O$_7$~\cite{moon2013}. In an ensuing series of papers~\cite{herbut2014, janssen2015a, janssen2015b}, we have argued, on the other hand, that the expansion around the upper critical dimension, and also the strict large-$N$ limit itself, might possibly even qualitatively fail to describe the ground-state behavior in the physical situation for $d=3$ and $N=1$. From a renormalization-group (RG) perspective, the chief culprit is the negligence of those short-range components of the Coulomb interaction, which one is tempted to discard as irrelevant at large $N$ or small $\epsilon=4-d$, which however may become important away from these limits. By employing a simple one-loop analysis for fixed $N=1$, we found the Abrikosov--Beneslavskii scale-invariant ground state to be unstable in $d=3$, indicated by a runaway flow of short-range interactions~\cite{herbut2014}. The situation is similar to $(2+1)$-dimensional relativistic quantum electrodynamics (QED$_{2+1}$), which has a scale-invariant (``conformal'') ground state at large number of fermions $N$, but is believed to suffer from a quantum phase transition towards a symmetry-broken ground state if $N$ drops below a critical number $\Nc$~\cite{appelquist1988, fischer2004, braun2014, dipietro2016, janssen2016, herbut2016}. In a calculation that resembles the inaugural work in QED$_{2+1}$, we showed that an analogous critical fermion number $\Nc$ exists in the nonrelativistic system with QBT in three spatial dimensions, by deriving and exploiting the nonperturbative solution of the Dyson-Schwinger equations within the $1/N$ expansion~\cite{janssen2015b}.

\begin{figure}[t]
\includegraphics[scale=1.1]{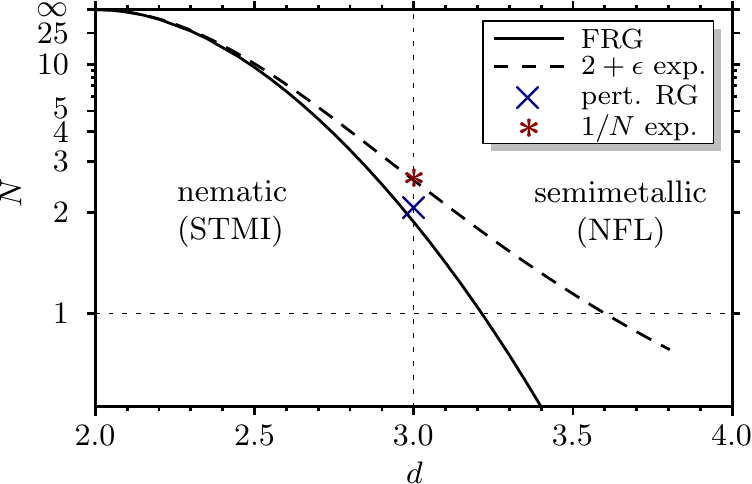}
\caption{Phase diagram of electronic systems with $N$ quadratic Fermi nodes in $2<d<4$ spatial dimensions from $2+\epsilon$ expansion (Sec.~\ref{sec:2+epsilon-expansion}), perturbative RG in fixed dimension (Sec.~\ref{sec:perturbative-expansion}), and functional RG (Sec.~\ref{sec:frg}). For comparison: $1/N$-expansion result from Ref.~\cite{janssen2015b}.
For visualization purposes, the vertical axis has been rescaled by $N/(1+N)$.
For $N<\Nc$ the system is unstable towards a nematic ground state with full, but anisotropic gap. For the systems with the band structure equivalent to that of HgTe, this state corresponds to a strong topological Mott insulator (STMI). For $N>N_c$ the systems remains semimetallic, but exhibits unusual exponents in various observables---a non-Fermi liquid state (NFL)~\cite{moon2013}. All these results place the physical situation for $N=1$ and $d=3$ (dotted lines) on the Mott insulating side of the transition.}
\label{fig:d-vs-Nfc}
\end{figure}

In the present work we add to the picture the results of various complementary approaches to the problem. We first demonstrate that above and near two spatial dimensions, the instability of the NFL state occurs at a large value of $N$. This allows us to prove the existence of a phase boundary in the $d$-$N$ plane, and to calculate its shape to leading order in $\epsilon = d-2$. Within this limit, we also show that the ground state for $N < \Nc$ is a \textit{nematic} insulator with spontaneously broken rotational symmetry and full, but anisotropic gap. We furthermore revisit the expansion near the upper critical dimension with control parameter $\epsilon = 4-d$ by formulating the corresponding Gross-Neveu-Yukawa theory, and establish the existence of another, quantum critical fixed point (QCP), alongside the Abrikosov-Beneslavskii NFL fixed point. The existence of this QCP turns out to be responsible for the nematic instability.
We demonstrate that both fixed points, and their nontrivial interplay as a function of $N$, can also be assessed in a simple perturbative expansion in fixed dimension $d=3$, thereby extending our previous results~\cite{herbut2014} to general fermion number $N$. If $N$ is lowered from infinity towards the physical $N = 1$, the QCP and the NFL fixed point approach each other in coupling space and eventually merge at some critical $\Nc$. For $N < \Nc$ they disappear into the complex-coupling plane, leaving behind the runaway flow of short-range couplings.
Computing the precise value for $\Nc$ in $d=3$ beyond simple approximations is, of course, a challenging task. In order to gain yet another estimate we employ the functional renormalization group (FRG) in the so-called dynamical bosonization scheme. Although this scheme \textit{a priori} lacks an obvious control parameter, we discover \textit{a posteriori} that our FRG predictions coincide precisely with the results from the $\epsilon$ expansions both near two and near four dimensions. For general $2<d<4$, it smoothly interpolates between these two perturbatively accessible limits~\cite{janssen2014}. This leads to a phase diagram in the $d$-$N$ plane with the insulating nematic state at small $N$ and/or near $d=2$, and the scale-invariant non-Fermi liquid state at large $N$ and/or near $d=4$; see Fig.~\ref{fig:d-vs-Nfc}.
In full agreement with our previous results, we find all these approaches to place the physical situation for $d=3$ and $N=1$ on the insulating side of the transition.
The resulting nematic ground state can be understood as if the material were under, in this case ``dynamically generated'', uniaxial strain. In the systems with the band structure equivalent to that of HgTe or $\alpha$-Sn, it corresponds to a strong topological insulator with dynamically generated band gap---a strong topological \textit{Mott} insulator.

The paper is organized as follows: We describe our model in Sec.~\ref{sec:model}. In Secs.~\ref{sec:2+epsilon-expansion} and \ref{sec:perturbative-expansion} we derive and discuss the $(2+\epsilon)$-expansion results and the results from perturbative expansion in fixed $d=3$. The properties of the nematic QCP are examined within an effective Gross-Neveu-Yukawa theory in Sec.~\ref{sec:4-epsilon-expansion}, within the expansion around the upper critical dimension $d=4$. Sec.~\ref{sec:frg} discusses our FRG approach to the problem. Concluding remarks are given in Sec.~\ref{sec:conclusions}.

\section{Model} \label{sec:model}

\begin{figure}[t]
\includegraphics[scale=0.92]{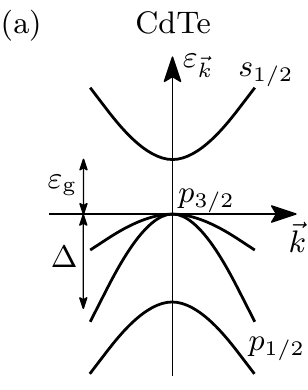}\hfill
\includegraphics[scale=0.92]{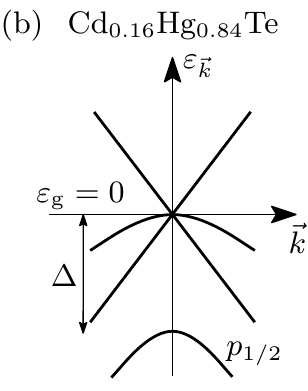}\hfill
\includegraphics[scale=0.92]{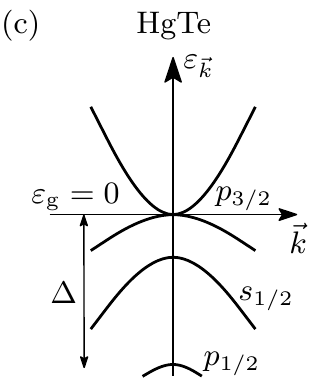}
\caption{Schematic electron band structure of II-VI zinc blende crystals near $\Gamma$ point. In the finite-gap semiconductor CdTe (a) the spin-orbit coupling splits the three $p$-type bands into a lower-lying $j=1/2$ band ($p_{1/2}$) and an energetically higher multiplet with total angular momentum $j=3/2$ ($p_{3/2}$).
In the solid solution Cd$_{1-x}$Hg$_{x}$Te spin-orbit coupling increases for increasing $x$. The $p_{3/2}$ multiplet is therewith shifted towards the conduction $s$ band, culminating in a linear band crossing at $x \approx 0.84$ (b). For even larger spin-orbit coupling the $s$ band exchanges roles with one of the $p_{3/2}$ bands (``band inversion''), and the band gap consequently remains identically zero, with the conduction and valence bands touching \textit{quadratically} at the Fermi level. This is the situation in the gapless semiconductor HgTe (c).}
\label{fig:electron-structure}
\end{figure}

Consider binary II-VI compounds crystallizing in the zinc blende structure, such as CdTe and HgTe.
CdTe is a semiconductor with a direct band gap of $\varepsilon_\text{g} \simeq 1.6\,\text{eV}$ at the center $\Gamma$ of the Brillouin zone~\cite{tsidilkovski1997}. In the nonrelativistic limit the valence-band upper edge at crystal momentum $\vec k=0$ consists of six degenerate $p$ states with orbital angular momentum quantum number $\ell=1$. Spin-orbit coupling partially lifts this degeneracy, giving rise to an energetically lower-lying doublet with total angular momentum $j=1/2$ (``$p_{1/2}$'' or ``$\Gamma_7$'' states) and an energetically higher-lying quadruplet with total angular momentum $j=3/2$ (``$p_{3/2}$'' or ``$\Gamma_8$'' states). The $p_{3/2}$ states at $\vec k = 0$ form the upper edge of the heavy- and light-hole bands. They are shifted by the spin-orbit interaction towards the energetically lowest conduction-band states which originate from the $s$ states of the next shell (``$s_{1/2}$'' or ``$\Gamma_6$'' states), thereby reducing the band gap; see Fig.~\ref{fig:electron-structure}(a).
Experimentally, the size of the band gap can be tuned by gradually substituting Cd atoms by Hg atoms in the solid solution Cd$_{1-x}$Hg$_x$Te. While Cd and Hg have the same valence shell electron configuration, the main difference is the influence of the relativistic effects due to the higher nuclear charge of Hg. Upon increasing $x$ in Cd$_{1-x}$Hg$_x$Te the gap between the $s_{1/2}$ electron band and the $p_{3/2}$ light- and heavy-hole bands shrinks; simultaneously, the effective masses of the electron band and the light-hole band decrease with increasing $x$. Eventually, at $x \approx 0.84$, the bands touch and the dispersion of both the electron band and the light-hole band becomes linear. By contrast, the effective mass of the heavy-hole band remains finite; see Fig.~\ref{fig:electron-structure}(b). For $x > 0.84$ the $s_{1/2}$ electron states drop below the $p_{3/2}$ multiplet and the curvature of the $s$-type band turns negative: it converts into a valence band. At the same time the curvature of the $p_{3/2}$ light-hole band becomes positive, thus now representing a conduction band; see Fig.~\ref{fig:electron-structure}(c). The band structure is \emph{inverted} and exhibits nontrivial topology with protected Dirac surface states~\cite{fu2007}. The system, however, is not (yet) a topological insulator: The degeneracy of the bulk $p_{3/2}$ states at the $\Gamma$ point is protected by the crystal symmetry. The band gap is therefore identically zero as long as the (discrete) rotational symmetry is not explicitly (e.g., by external strain) or spontaneously (e.g., by interactions) broken. Moreover, in the undoped system, the Fermi level is locked right at the QBT.
The touching point is fourfold degenerate and consists of states with total angular momentum quantum number $j=3/2$. Consequently, the effective low-energy Hamiltonian can be written in terms of $4\times 4$ angular momentum representation matrices $J_x, J_y, J_z \in \mathbb C^{4 \times 4}$.
Taking the crystal's cubic symmetry into account, the only possible invariants that can be constructed out of these matrices and which are quadratic in momentum $\vec k$ are given by
\begin{equation}
 \mathbbm 1_{4 \times 4}, \qquad
 (\vec k \cdot \vec J)^2, \qquad
 k_x^2 J_x^2 + k_y^2 J_y^2 + k_z^2 J_z^2.
\end{equation}
In the above, the first two invariants respect the full spherical $\mathrm O(3)$ symmetry, while the third one breaks the rotational symmetry down to the discrete cubic $C_4 \times C_4$ symmetry. Experimentally, the degree of nonsphericity is usually relatively small~\cite{tsidilkovski1997}. In fact, the full spherical symmetry is expected to be \emph{emergent} at low energies~\cite{abrikosov1974}.

A low-energy model for the Fermi systems with quadratic band touching is therefore given by the spherically symmetric Luttinger Hamiltonian~\cite{luttinger1956}
\begin{equation}
 H_0(\vec k) = \frac{1}{2m_0} \left[\left(\alpha_1 + \frac{5}{2} \alpha_2 \right) p^2 \mathbbm 1_{4 \times 4} - 2\alpha_2 (\vec k \cdot \vec J)^2 \right],
\end{equation}
with electron mass $m_0$ and phenomenological parameters $\alpha_1$ and $\alpha_2$. The spectrum of $H_0$ is
\begin{equation}
 \varepsilon_{\vec k} = (\alpha_1 \pm 2\alpha_2) \frac{{\vec k}^2}{2m_0},
\end{equation}
and describes a QBT at $\vec k = 0$ if $|\alpha_1| < 2|\alpha_2|$.
The parameters $\alpha_1$ and $\alpha_2$ can be extracted experimentally by fitting, for instance, magnetoabsorbtion measurements to the predictions of the Luttinger model, yielding
\begin{align}
 \alpha_1 & \simeq 12.8, &
 \alpha_2 & \simeq 8.4,
\end{align}
at temperature $T=4.4\,\mathrm K$~\cite{guldner1973}. Such values correspond to effective masses of the electron and hole bands near the $\Gamma$ point as $m_{\mathrm e} \simeq \frac{m_0}{30}$ and $m_{\mathrm h} \simeq \frac{m_0}{4}$. The electron effective mass is significantly smaller than the hole effective mass. This is because at the band inversion point the former vanishes while the latter remains finite, see Fig.~\ref{fig:electron-structure}(b).
Nevertheless, at low temperatures and in pure samples, it is theoretically expected that electron and hole effective masses renormalize and eventually approach a common value towards $T \to 0$---emergent particle-hole symmetry~\cite{moon2013, boettcher2016}. Plasma reflection measurements in p-doped HgTe indeed show a systematic decrease of the hole effective mass with decreasing charge carrier concentration, and this has been attributed to the effect of electron-electron interactions~\cite{ivanovomskii1983}. To the best of our knowledge, however, emergent particle-hole symmetry has not yet been experimentally verified so far.

To simplify the problem, we here assume both particle-hole and spherical symmetry from the outset. The intricate effects of symmetry-breaking terms will be discussed in a separate paper \cite{boettcher2}. We thus set $\alpha_1 \equiv 0$ and absorb $2\alpha_2$ in the definition of the effective mass $m \equiv m_0/(2\alpha_2)$. The spectrum hence simply becomes $\varepsilon_{\vec k} = \pm k^2/(2m)$.
The Luttinger Hamiltonian can then be written in a form that allows its immediate generalization to arbitrary dimension $d$~\cite{janssen2015a}:
\begin{align} \label{eq:hamiltonian}
 H_0(\vec k) = \sum_{a=1}^{(\frac{d}{2}+1)(d-1)} d_a(\vec k) \gamma_a,
\end{align}
with Hermitian Dirac matrices $\gamma_a$, $a=1,\dots,(d/2+1)(d-1)$, satisfying the Clifford algebra $\{\gamma_a, \gamma_b\} = 2\delta_{ab}$. They have dimension $d_\gamma = 2^{\lfloor(d+2)(d-1)/4\rfloor}$ with $\lfloor \, \cdot \, \rfloor$ symbolizing the floor function. The functions $d_a(\vec k)$ denote real hyperspherical harmonics for the angular momentum of two on the $(d-1)$-sphere:
\begin{align} \label{eq:hyperspherical}
 d_a(\vec k) = \sqrt{\frac{d}{2(d-1)}} \sum_{i,j=1}^{d} k_i \Lambda_{a,ij} k_j,
\end{align}
with the generalized real Gell-Mann matrices $\Lambda_{a}$ as constructed in Ref.~\cite{janssen2015a}. 
In $d=3$, they read
\begin{gather}
 \Lambda_1 = 
 \begin{pmatrix}
  1 & 0 & 0 \\
  0 & -1 & 0 \\
  0 & 0 & 0
 \end{pmatrix}, \qquad
 \Lambda_2 = 
 \begin{pmatrix}
  0 & 1 & 0 \\
  1 & 0 & 0\\
  0 & 0 & 0
 \end{pmatrix}, \displaybreak[0] \nonumber\\
 \Lambda_3 = 
 \begin{pmatrix}
  0 & 0 & 1 \\
  0 & 0 & 0 \\
  1 & 0 & 0
 \end{pmatrix}, \qquad
 \Lambda_4 = 
 \begin{pmatrix}
  0 & 0 & 0 \\
  0 & 0 & 1 \\
  0 & 1 & 0
 \end{pmatrix}, \displaybreak[0] \nonumber\\
 \Lambda_5 = 
 \frac{1}{\sqrt{3}}
 \begin{pmatrix}
  -1 & 0 & 0\\
  0 & -1 & 0\\
  0 & 0 & 2
 \end{pmatrix}.
\end{gather}
From this, we obtain, for $d=3$, $d_1+\ii d_2 = \frac{\sqrt{3}}{2} k^2 \ee^{2\ii \varphi}\sin^2\vartheta$, $d_3+\ii d_4 = \frac{\sqrt{3}}{2} k^2 \ee^{\ii\varphi} \sin 2\vartheta$, and $d_5 = \frac{1}{2}k^2(3\cos^2\vartheta - 1)$, with $\vartheta$ and $\varphi$ as spherical angles in $\vec k$~space. The five Dirac matrices $\gamma_1,\dots,\gamma_5$ in this case have dimension $d_\gamma=4$. This form of the three-dimensional Luttinger Hamiltonian was previously put forward in the context of finite-gap semiconductors with zinc blende structure in Ref.~\cite{murakami2004}. Setting $d=2$, we recover the known two-dimensional QBT Hamiltonian~\cite{sun2009, vafek2010, dora}. For $d=4$, Eqs.~\eqref{eq:hamiltonian} and \eqref{eq:hyperspherical} become equivalent to the four-dimensional QBT Hamiltonian proposed by Abrikosov~\cite{abrikosov1974}.

In the continuum limit, the noninteracting Lagrangian that describes the electronic system with $N$ independent quadratic nodes at the Fermi level is then given by
\begin{equation} \label{eq:L-0}
 L_0 = \psi_i^\dagger \left[\partial_\tau + H_0(-\ii\nabla)\right]\psi_i, \qquad i=1,\dots,N.
\end{equation}
Each Grassmann field $\psi_1, \dots,\psi_N$ has $d_\gamma$ components. $\tau$ denotes imaginary time.
In Eq.~\eqref{eq:L-0} and from now on, where unambiguous, we assume summation convention over repeated indices. The physical situation that describes HgTe and $\alpha$-Sn, as well as the pyrochlore iridates and half-Heusler compounds, is given by $N=1$. For generality and computational clarity, however, we allow an arbitrary number $N$ of fermion species, i.e., Fermi nodes.

As the Coulomb interaction is only partially screened~\cite{janssen2015b}, its effects are crucial. We take them into account by introducing a scalar field $a$ which mediates the long-range interaction via
\begin{equation}
 L_a = \frac{1}{2} (\nabla a)^2 + \ii e a \psi_i^\dagger \psi_i.
\end{equation}
Here, $e$ is the effective charge, into which all numerical constants of the system have been absorbed, $e^2 = 2m e_0^2 / (4\pi \hbar^2 \kappa)$. $e_0$ is the electron charge and $\kappa$ denotes the ``background'' dielectric constant arising from energy bands away from the Fermi level~\cite{halperin1968}. Integrating out the Coulomb field $a$ in the partition function would generate a nonlocal density-density interaction $\propto 1/k^2$ in Fourier space, which is $\propto 1/r^{d-2}$ in real space, and thus of the usual form in $d=3$.

\section{$2+\epsilon$ expansion} \label{sec:2+epsilon-expansion}

In a RG treatment, new effective interactions which are not present microscopically may be generated by loop corrections. Most of them are power-counting irrelevant and can be safely neglected. Local four-fermion interactions, however, are marginal in $d=2$ and could thus become relevant at an interacting fixed point in $d=2+\epsilon$. We therefore have to take them into account as well.
In fact, the diagrams shown in Fig.~\ref{fig:box-charge} generate a local four-fermion term of the form $\propto e^4 (\psi^\dagger_i \gamma_a \psi_i)^2$ from the long-range interaction. Once generated, this term generates further contact interactions. At one-loop order, we find that all possible four-fermion interactions can be written as a linear combination of the following three basis terms:
\begin{figure}[t]
 \includegraphics{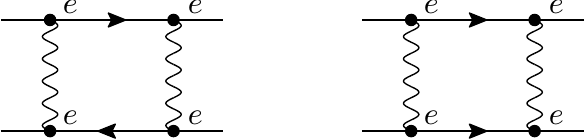}
 \caption{Feynman diagrams that generate local four-fermion interactions from the long-range Coulomb interactions at one-loop order. Straight (wiggly) inner lines correspond to fermion (Coulomb) propagators.}
 \label{fig:box-charge}
\end{figure}
\begin{equation} \label{eq:L-psi}
 L_{\psi} = g_1 (\psi^\dagger_i \psi_i)^2 + g_2 (\psi^\dagger_i \gamma_a \psi_i)^2 + g_3 (\psi^\dagger_i \gamma_{ab} \psi_i)^2,
\end{equation}
with $\gamma_{ab} = \frac{\ii}{2}[\gamma_a,\gamma_b]$ and couplings $g_\alpha$, $\alpha=1,2,3$.

The four-fermion couplings have mass dimension $[g_\alpha] = 2-d$, suggesting that an $\epsilon$ expansion around $d=2$ may be feasible. The charge $e$, however, has dimension $[e^2] = 4-d$ and is thus strongly RG relevant towards the infrared. In order to gain full control over the perturbative expansion, we have to take the limit of both small $\epsilon$ and large $N$. In this double-expansion limit proper fixed points become weakly coupled in all interaction channels. Fortunately, as we shall see below, the fixed-point annihilation we are after automatically happens at large $N$ as long as $\epsilon$ is small. The scenario is therefore under perturbative control with a single control parameter $\epsilon=d-2$.

The RG flow in the full theory space, given by
\begin{equation} \label{eq:L-2+eps}
 L = L_0 + L_a + L_\psi,
\end{equation}
is obtained by integrating out a thin momentum shell from the ultraviolet cutoff $\Lambda$ to $\Lambda/b$ with $b>1$. At one-loop order, we find the flow equations
\begin{align}
 \label{eq:beta-e2}
 \frac{d e^2}{d\ln b} & = (2+z-d - \eta_a) e^2, \\
 \label{eq:beta-g1}
 \frac{d g_1}{d \ln b} & = (z-d)g_1 - (e^2 + 2 g_1) g_2 - 24 g_3^2, \\
 \label{eq:beta-g2}
 \frac{d g_2}{d \ln b} & = (z-d)g_2 + \frac{4(e^2+2g_1) g_2}{5}  - \frac{(e^2+2g_1)^2}{20}
 \nonumber \\ & \quad
 - \frac{37+16N}{5} g_2^2 + \frac{112}{5} g_2 g_3 - \frac{136}{5} g_3^2, \\
 \label{eq:beta-g3}
 \frac{d g_3}{d \ln b} & = (z-d)g_3 - \frac{1}{5}(e^2+2g_1)g_3 + g_2^2 - 6 g_2 g_3
 \nonumber \\ & \quad
 + \frac{4(11-4N)}{5}g_3^2
\end{align}
with the anomalous dimension of the Coulomb field $\eta_a$ and dynamical exponent $z$ as
\begin{align} \label{eq:eta-z}
 \eta_a & =  N e^2, &
 z & = 2 - \frac{4}{15} e^2.
\end{align}
In order to arrive at Eqs.~\eqref{eq:beta-e2}--\eqref{eq:eta-z} we have rescaled the couplings as $e^2 \Lambda^{d+\eta_a-z-2} S^d/(2\pi)^d \mapsto e^2$ and $g_\alpha \Lambda^{d-z} S^d/(2\pi)^d \mapsto g_\alpha$ with $S^d$ as the surface area of the $(d-1)$-hypersphere in $d$ dimensions. We have performed the angular integration directly in $d=3$, and have employed the $4\times 4$ representation of the Dirac matrices $\gamma_a$, with $a=1,\dots,5$. The dimension of the couplings $e^2$ and $g_\alpha$ is counted in general dimension $d$.
These flow equations generalize those previously published~\cite{herbut2014} to general fermion number $N> 1$. To see that they coincide with the latter in the limit $N \searrow 1$ we note that in this limit one of the terms in Eq.~\eqref{eq:L-psi} can be eliminated in favor of the other two by making use of the Fierz identity~\cite{hjr, herbut2014}
\begin{align}
 \sum_{a,b=1}^{5} (\psi^\dagger \gamma_{ab} \psi)^2 = - 10 (\psi^\dagger \psi)^2 - 2 \sum_{a=1}^{5}(\psi^\dagger \gamma_a \psi)^2.
\end{align}
For $N=1$ we can thus rewrite Eq.~\eqref{eq:L-psi} as
\begin{align}
 L_\psi^{N=1} = (g_1 - 10 g_3)(\psi^\dagger\psi)^2 + (g_2 - 2 g_3) \sum_{a=1}^5 (\psi^\dagger \gamma_a \psi)^2,
\end{align}
which is the form employed in Ref.~\cite{herbut2014}. If we now shift the couplings in Eqs.~\eqref{eq:beta-g1}--\eqref{eq:beta-g3} appropriately,
\begin{align}
 g_1 - 10 g_3 & \mapsto g_1, &
 g_2 - 2 g_3 & \mapsto g_2,
\end{align}
we find that the flow equations for the shifted couplings $g_1$ and $g_2$ indeed become independent of the third coupling $g_3$ when $N=1$:
\begin{align} \label{eq:beta-g1-N1}
 \frac{d g_1}{d \ln b} & = (z-d) g_1 - (e^2 + 2 g_1) g_2 - 10 g_2^2, \\
 \frac{d g_2}{d \ln b} & = (z-d) g_2 - \frac{(e^2 + 2 g_1)^2}{20} + \frac{4(e^2 + 2g_1) g_2}{5}
 \nonumber \\ &\quad 
 -\frac{63}{5} g_2^2.
 \label{eq:beta-g2-N1}
\end{align}
Moreover, Eqs.~\eqref{eq:beta-g1-N1} and \eqref{eq:beta-g2-N1} in conjunction with Eqs.~\eqref{eq:beta-e2} and \eqref{eq:eta-z} for $N=1$ are evidently equivalent to the flow equations published in Ref.~\cite{herbut2014}. Our results for general $N$ are therefore continuously connected to the $N=1$ case, the presence of a third independent coupling for $N>1$ notwithstanding.

Some comments on the structure of the flow equations are in order:
\begin{enumerate}[(1)]
\item If we start the RG for small (but finite) initial charge $e^2>0$ and vanishing contact interactions $g_\alpha=0$ (as relevant for HgTe and $\alpha$-Sn), $e^2$ will flow to larger values towards the infrared until the anomalous dimension $\eta_a$ becomes of the order $2+z-d$ when the flow of $e^2$ slows down and eventually stops for $\eta_a = 2+z-d$ [see Eq.~\eqref{eq:beta-e2}].
\item The beta function for the charge $e^2$ has an especially simple form \cite{herbut2001}; in particular, there is no vertex correction $\propto e^2 g_\alpha$ in $(d e^2)/(d\ln b)$. That this happens here at one loop is not a coincidence, but basically a consequence of the Ward identity associated to the gauge symmetry $\psi \mapsto \ee^{\ii e\lambda(\tau)}\psi$, $a \mapsto a - \partial_\tau \lambda$. We therefore expect the form of Eq.~\eqref{eq:beta-e2} [but not Eq.~\eqref{eq:eta-z}] to hold at arbitrary loop order. In this way we obtain an \emph{exact} relation for the Coulomb anomalous dimension $\eta_{a}$ at a putative charged fixed point,
\begin{equation} \label{eq:eta-a-exact}
 \eta_{a} = 2 + z - d,
\end{equation}
with $z$ being the (presumably nontrivial) dynamical exponent at the fixed point.
This resembles the analogous situation in the Abelian Higgs model and in QED$_{2+1}$, where similar exact relations for the gauge anomalous dimensions are known~\cite{herbut1996, janssen2016}.
The form of the (marginally) screened Coulomb potential at a charged fixed point is therefore $V(r) \propto 1/r^{z}$ exactly. This is in agreement with the large-$N$ result of Ref.~\cite{janssen2015b}.
\item Using the exact relation~\eqref{eq:eta-a-exact} together with the one-loop formulae for $\eta_a$ and $z$ in Eq.~\eqref{eq:eta-z} we obtain the fixed-point value for the charge: $e^2_* = (4 - d)/N + \mathcal O(1/N^2)$, which is under perturbative control in the limit of large $N$. In this limit we find
\begin{align}
 \eta_{a} & = (4-d)\left[1-\frac{4}{15N} + \mathcal{O}(1/N^2)\right],
 \label{eq:eta-a-largeN} \\
 z & = 2 - \frac{4(4-d)}{15N} + \mathcal O(1/N^2),
 \label{eq:z-largeN}
\end{align}
at a putative charged fixed point, in agreement with Ref.~\cite{moon2013}.
\end{enumerate}

In the following we describe the fixed-point structure in the double-expansion limit $1/N \ll \epsilon \ll 1$ with $\epsilon = d-2$. It will prove convenient to consider a finite and fixed, but arbitrary, product $n \equiv N\epsilon^2$. In the limit of small $\epsilon$ with fixed $n$, the flow equations decouple and no contact interactions $\propto g_1$ and $\propto g_3$ will be generated by the charge, if absent initially. The coupling $g_2$, however, will be generated according to the flow equation
\begin{equation} \label{eq:beta-g2-2+eps}
 \frac{d g_2}{d\ln b} = -\epsilon g_2 -\frac{16 g_2^2}{5} - \frac{n e^4}{20 \epsilon^2},
\end{equation}
where we have rescaled $g_\alpha \mapsto g_\alpha/N$ for convenience. The charge sector has a stable fixed point at $e_*^2 = 2\epsilon^2/n + \mathcal O(\epsilon^4)$. With this value for the charge, the above flow equation has fixed points at
\begin{equation} \label{eq:fp-g2-2+eps}
 g_{2*}^{\pm} = - \frac{5\epsilon}{32}\left(1 \pm \sqrt{1-\frac{64/25}{n}}\right) + \mathcal O(\epsilon^2).
\end{equation}
They are located at real values of the coupling if and only if $n \geq 64/25$, i.e.,
\begin{equation} \label{eq:criticalN-2+eps}
 N \geq \Nc(\epsilon) = \frac{64}{25\epsilon^2} + \mathcal O(1/\epsilon),
\end{equation}
to the leading order in $\epsilon=d-2$.

\begin{figure}[t]
 \includegraphics{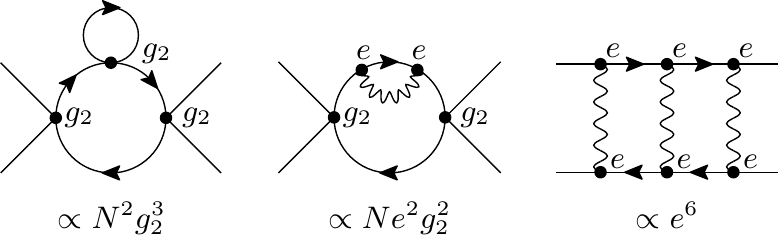}
 \caption{Two-loop Feynman diagrams contributing to the flow of $g_2$ at large $N$. From the number of closed fermion loops we can deduce the scaling of these terms with $N$ to leading order in $1/N$, which after the rescaling [as below Eq.~\eqref{eq:beta-g2-2+eps}] yield the contributions as displayed in Eq.~\eqref{eq:two-loop}.}
 \label{fig:two-loop}
\end{figure}

Higher-order loop corrections indeed contribute only to subleading order $\mathcal O(1/\epsilon)$ to $\Nc$, as anticipated in the above equation. This can be argued as follows: To two-loop order, there are three classes of diagrams that contribute to the beta function of $g_2$ as $\propto g_2^3$, $\propto e^2 g_2^2$, and $\propto e^6$, respectively. Examples for each class are given in Fig.~\ref{fig:two-loop}.  Without explicitly evaluating the diagrams, we can deduce their leading-order scaling with $N$ by counting the number of closed fermion loops in each diagram. 
The examples in Fig.~\ref{fig:two-loop} are representatives of those diagrams that contribute to the leading order for large $N$.
From this we obtain the form of the two-loop corrections to the flow of $g_2$ within our double-expansion limit as [after the rescaling as below Eq.~\eqref{eq:beta-g2-2+eps}]
\begin{align}\label{eq:two-loop}
 \left. \frac{d g_2}{d \ln b} \right\rvert_\text{two loop} = c_1 g_2^3 + c_2 e^2 g_2^2 + \frac{c_3 n e^6}{\epsilon^2},
\end{align}
with \emph{fixed} coefficients $c_{1,2,3}$. These terms evidently contribute only to order $\mathcal O(\epsilon^3)$ to the flow of $g_2$ when $g_{2} = \mathcal O(\epsilon)$ and $e^2 = \mathcal O(\epsilon)$, while the tree-level and one-loop terms given in Eq.~\eqref{eq:beta-g2-2+eps} are of order $\mathcal O(\epsilon^2)$. We expect a similar hierarchy to hold also beyond the two-loop order. Consequently, the one-loop value for $\Nc$ as displayed in Eq.~\eqref{eq:criticalN-2+eps} is the correct leading-order value for small $\epsilon$.

The presence (absence) of the two charged fixed points for $N$ above (below) $\Nc$ has striking implications for the structure of the RG flow. Consider first $N > \Nc$: The fixed point at $g_{2*}^-$ is fully infrared attractive, representing a scale-invariant phase with gapless fermions but nontrivial exponents as given in Eqs.~\eqref{eq:eta-a-largeN} and \eqref{eq:z-largeN}. The weakly-interacting regime with $e^2 \ll 1$ and $g_\alpha = 0$ indeed lies in this fixed point's basin of attraction. The fixed point represents a conformal phase and is nothing but the Abrikosov-Beneslavskii non-Fermi-liquid fixed point previously found at large $N$~\cite{abrikosov1974, moon2013}.
The fixed point at $g_{2*}^+$, on the other hand, is a quantum critical point with precisely one RG relevant direction. It was found earlier within a simple perturbative RG analysis for $N=1$~\cite{herbut2014}. The strongly-interacting regime for $g_2 < g_{2*}^+ < 0$ is no longer in the basin of attraction of the Abrikosov-Beneslavskii NFL fixed point, but exhibits an instability towards divergent coupling $g_2 \to -\infty$ at finite RG scale. The transition is governed by the QCP at $g_{2*}^+ $.
In order to elucidate the nature of the instability and the corresponding infrared phase we add to the Lagrangian various types of infinitesimally small symmetry-breaking bilinears
\begin{align} \label{eq:L-Delta}
 L_\Delta & =
 \Delta_\text{nem} \psi^\dagger_i \gamma_5 \psi_i
 + \Delta_\text{sc} \psi^\dagger_i \gamma_{45} \psi_i^*
 + \Delta_\text{sc}^* \psi^\mathrm{T}_i \gamma_{45} \psi_i
 \nonumber \\ &\quad
 + \Delta_\text{ch} \psi^\dagger_i \psi_i
 + \Delta_\text{mag} \psi^\dagger_i \gamma_{45} \psi_i,
\end{align}
where $\gamma_{45} \equiv \ii \gamma_4 \gamma_5$. The \emph{nematic} order parameter $\Delta_\text{nem}$ breaks the rotational symmetry and can be understood as originating from uniaxial strain in $z$ direction~\cite{roman1972, moon2013}. Finite $\Delta_\text{nem}$ opens a full, but anisotropic gap in the spectrum and converts the semimetal into a three-dimensional topological insulator~\cite{fu2007, bruene2011}. In the related uncharged system with the long-range Coulomb interaction neglected, nematic quantum criticality was previously extensively investigated by devising and exploiting the corresponding Gross-Neveu-Yukawa theory in $d=4-\epsilon$ dimensions~\cite{janssen2015a}.
The superconducting $s$-wave order parameter $\Delta_\text{sc}$, on the other hand, breaks $\mathrm U(1)$ charge symmetry. A corresponding superconducting quantum critical point for strong attractive contact interactions was also recently investigated~\cite{boettcher2016}. $\Delta_\text{ch}$ induces a finite charge density, and represents a finite chemical potential. Finally, $\Delta_\text{mag}$ breaks time reversal. The corresponding magnetic quantum critical point, which in the pyrochlore iridates governs a transition towards an all-in-all-out antiferromagnet, was previously studied at large $N$~\cite{savary2014}.
To the leading order in the present $(\epsilon, 1/N)$ double expansion with fixed $n = N \epsilon^2$ the flow of these ``mass parameters'' reads
\begin{align}
 \frac{d \Delta_\text{nem}}{d \ln b} & = \left(z - \frac{16 g_2}{5}  + \frac{2e^2}{5}  \right) \Delta_\text{nem} + \mathcal O(\Delta_\alpha^2), \displaybreak[0] 
 \label{eq:beta-Delta-A} \\
 \frac{d \Delta_\text{sc}}{d \ln b} & = \left(z - \frac{e^2}{2} \right) \Delta_\text{sc} + \mathcal O(\Delta_\alpha^2), \displaybreak[0] 
 \label{eq:beta-Delta-B} \\
 \frac{d \Delta_\text{ch}}{d \ln b} & = z\, \Delta_\text{ch} + \mathcal O(\Delta_\alpha^2), \displaybreak[0] 
 \label{eq:beta-Delta-C} \\
 \frac{d \Delta_\text{mag}}{d \ln b} & = \left(z + \frac{e^2}{5} \right) \Delta_\text{mag} + \mathcal O(\Delta_\alpha^2),
 \label{eq:beta-Delta-D}
\end{align}
where we have applied the same rescalings as below Eqs.~\eqref{eq:eta-z} and \eqref{eq:beta-g2-2+eps}. Near the QCP $(e^2_*, g_{2*}^+)$ the free energy density has a scaling form~\cite{herbut2007}:
\begin{align}
 f(\delta g, \Delta_\alpha) = |\delta g|^{(d+z)/y} \,F_\alpha^\pm \!\left(\frac{\Delta_\alpha}{\lvert \delta g \rvert^{x_\alpha/y}}\right),
\end{align}
where $\delta g \equiv g_2 - g_{2*}$ defines the distance to criticality and $F_\alpha^\pm$ is a scaling function. The exponents $x_\alpha$ and $y$ are given by the linearized flow of $\Delta_\alpha$ and $\delta g$,
\begin{align}
 \frac{d \Delta_\alpha}{d \ln b} & = x_\alpha \Delta_\alpha + \mathcal O(\Delta^2), &
 \frac{d \delta g}{d \ln b} & = y \, \delta g + \mathcal O(\delta g^2),
\end{align}
with $\alpha \in \{\mathrm{nem}, \mathrm{sc}, \mathrm{ch}, \mathrm{mag}\}$. The corresponding susceptibilities $\chi_\alpha$ therefore scale as
\begin{align}
 \chi_\alpha = - \frac{\partial^2 f}{\partial \Delta_\alpha^2} \propto |\delta g|^{-\gamma_\alpha} \quad \text{with} \quad \gamma_\alpha = \frac{2x_\alpha-d-z}{y}
\end{align}
near the QCP. $\chi_\alpha$ diverges if $\gamma_\alpha > 0$. From Eqs.~\eqref{eq:fp-g2-2+eps} and \eqref{eq:beta-Delta-A}--\eqref{eq:beta-Delta-D} we find for $n > 64/25$ to the leading order
\begin{align}
 \gamma_\mathrm{nem}/\nu & = \epsilon \sqrt{1-\frac{64/25}{n}} + \mathcal O(\epsilon^2), \\
 \gamma_\mathrm{sc}/\nu & = -\epsilon + \mathcal O(\epsilon^2), \\
 \gamma_\mathrm{ch}/\nu & = -\epsilon + \mathcal O(\epsilon^2), \\
 \gamma_\mathrm{mag}/\nu & = -\epsilon + \mathcal O(\epsilon^2),
\end{align}
with the correlation-length exponent $\nu = 1/y = 1/[\epsilon \sqrt{1-64/(25n)}]$.
At the QCP, there is therefore a unique ordering tendency that corresponds to a positive susceptibility exponent: the nematic instability with order parameter $\langle \psi^\dagger_i \gamma_5 \psi_i \rangle$ and exponent $\gamma_\mathrm{nem} = 1 + \mathcal O(\epsilon)$. Note also that $\Delta_\text{ch}$ (chemical potential) couples to the conserved charge in the theory and as such did not receive perturbative corrections in Eq.~\eqref{eq:beta-Delta-C}.

\begin{figure}[t]
 \includegraphics{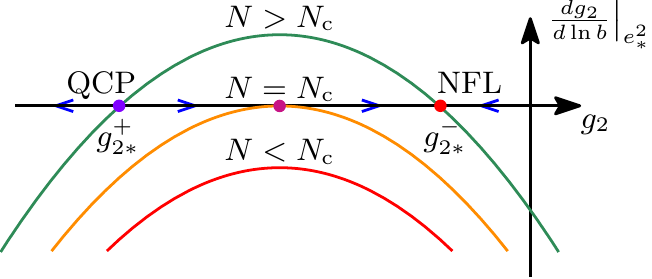}
 \caption{Schematic beta function for $g_2$ with $e^2$ at its infrared fixed-point value $e^2_*$. Blue arrows on horizontal axis indicate RG flow towards infrared. For $N \searrow \Nc$ the QCP and the NFL fixed point merge and annihilate for $N < \Nc$, leaving behind the runaway flow towards negative $g_2$.}
 \label{fig:beta-g2}
\end{figure}

Consider now $N$ near $\Nc$: The two fixed points at $g_{2*}^+$ and $g_{2*}^-$ then approach each other and eventually merge at $N \searrow \Nc$. Below $\Nc$, they disappear into the complex-coupling plane. The beta function $(d g_2)/(d \ln b)$ [Eq.~\eqref{eq:beta-g2-2+eps}] for $e^2 = e^2_*$ is then always negative, see Fig.~\ref{fig:beta-g2}.
The flow for $N < \Nc$ is therefore towards divergent negative $g_2$ for \emph{all} ultraviolet starting values for $g_2$, i.e., even in the weakly-interacting limit with $\left.g_2 \right\rvert_\text{UV} \approx 0$, relevant for HgTe and $\alpha$-Sn.

When the fixed-point annihilation takes place, the correlation-length exponent $\nu$ at the QCP diverges, which is consistent with the infinite-order transition at $\Nc$, see below. The susceptibility exponent $\gamma_\mathrm{nem}$, on the other hand, remains finite. Note that the structure of the flow diagram changes only locally near $g_{2*}^+ = g_{2*}^-$ when $N$ crosses $\Nc$, with the behavior away from this merging point remaining unchanged. By continuity, we therefore expect that the nature of the infrared phase for $N$ near and below $\Nc$ with small $\left.g_2 \right\rvert_\text{UV} \approx 0$ is the \emph{same} as the infrared phase for $N$ near and above $\Nc$ with $\left.g_2 \right\rvert_\text{UV} < g_{2*}^+$.

This way, we conclude that the electronic systems with quadratic Fermi nodes and (even weak) long-range Coulomb interaction is unstable towards the nematic ordering when $N < \Nc$, with $\Nc = 64/(25 \epsilon^2) + \mathcal O(1/\epsilon)$ to the leading order in the $2+\epsilon$ expansion. This is in agreement with the previous result for $N=1$~\cite{herbut2014}. Naive extrapolation of the critical fermion number to the physical dimension $d=3$ leads to $\Nc(d=3) \simeq 2.56$, and thus above the physical case for $N=1$. The low-$N$ system appears as if under, in this case dynamically generated, uniaxial strain, and represents a topological Mott insulator. 

Near and below $\Nc$, the RG flow effectively slows down at the merging point. Upon integrating the flow equations we find that the RG ``time'' $b$ it takes the flow of $g_2$ to diverge is
\begin{equation}
 b_0 = \exp\left( \frac{\pi/\epsilon}{\sqrt{\frac{\Nc}{N} - 1}} + \mathcal O[(\Nc/N-1)^0] \right).
\end{equation}
For $N \lesssim \Nc$, the dynamically generated gap $\Delta$ is hence exponentially suppressed,
\begin{equation} \label{eq:gap}
 \Delta \propto b_0^{-z}.
\end{equation}
The above scaling law has an essential singularity at $N \nearrow \Nc$, much like the thermal Berezinskii-Kosterlitz-Thouless transition~\cite{herbut2007}, and quite typical for the present type of conformal phase transition~\cite{kaplan2009}. The result is consistent with the form previously derived for the 3D QBT system by solving the Dyson-Schwinger equations directly in $d=3$ within the $1/N$ expansion~\cite{janssen2015b}. Here, this infinite-order transition as function of $N$ follows as a direct consequence of the fixed-point annihilation mechanism. Analogous scaling laws for $N$-dependent transitions are known to hold also for conformal phase transitions in QED$_3$~\cite{appelquist1988, janssen2016, herbut2016}, the Abelian Higgs model~\cite{halperin1974, herbut1996, herbut2007}, and many-flavor quantum chromodynamics~\cite{jaeckel2006}, and in all of these cases are due to an analogous mechanism.

\section{Perturbative RG in fixed $d=3$} \label{sec:perturbative-expansion}

Now that the \emph{existence} of a finite critical fermion number $\Nc$ in $d>2$ is established within a controlled $2+\epsilon$ expansion, a natural next step is to predict its \emph{value} in the physical case for $d=3$. This is a difficult strong-coupling problem. Similar to classical critical phenomena, a reliable theoretical estimate can only be obtained by employing and comparing various different approaches to the problem. While the large-$N$ theory in fixed $d=3$ has been devised recently~\cite{janssen2015b}, another simple approach is the perturbative renormalization group in fixed dimension. This is the subject of the present section, thereby generalizing the $N=1$ results of Ref.~\cite{herbut2014} to $N > 1$. Yet another approach to the problem will be employed in Sec.~\ref{sec:frg}.

The downside of this perturbative approach is the lack of a small control parameter, as the fixed-point annihilation will take place in a strong-coupling regime. Ignoring this reservation, we may evaluate the flow equations \eqref{eq:beta-e2}--\eqref{eq:eta-z} directly in $d=3$. Although the structure of the flow considerably gains in complexity as the different short-range couplings no longer decouple near $\Nc$, the physical conclusions drawn within the $2+\epsilon$ expansion entirely carry over to the present approach:

Small initial charge $e^2$ flows to strong coupling towards the infrared with the fixed-point value being $e^2_* = 15/(15 N + 4)$. At a putative charged fixed point the dynamical exponent is $z = 2 - 4/(15N+4)$ and the Coulomb anomalous dimension is $\eta_{a} = 15 N/(15N+4)$, in agreement with the exact relation, Eq.~\eqref{eq:eta-a-exact}.
For $N > \Nc$ with
\begin{align}
 \Nc = 2.095
\end{align}
we find an RG attractive NFL fixed point located at real couplings.
In the large-$N$ limit it is located at
\begin{equation} \label{eq:g1-g2-g3-NFL}
 \text{NFL:} \quad \left(g_{1*}, g_{2*}, g_{3*}\right) = \left(0, -\frac{1}{20N^2}, 0 \right) + \mathcal O(1/N^3).
\end{equation}
It governs the infrared behavior of the weakly-interacting theory with ultraviolet values $\left. g_{\alpha} \right|_\text{UV} \approx 0$, $\alpha = 1, 2, 3$, and $\left. e^2  \right|_\text{UV}> 0$.
There is also a QCP with one RG relevant direction. It is located at strong repulsive short-range coupling $g_{2*} < 0$ and weak $\lvert g_{{1,3}*} \rvert \ll |g_{2*}|$.
In the large-$N$ limit its fixed-point values read
\begin{equation} \label{eq:g1-g2-g3-QCP}
 \text{QCP:} \quad \left(g_{1*}, g_{2*}, g_{3*}\right) = \left(0, -\frac{5}{16N}, 0 \right) + \mathcal O(1/N^2).
\end{equation}
For $N \searrow \Nc$ the QCP and the NFL fixed point merge at
\begin{align} \label{eq:g1-g2-g3-fixed-d}
 \text{QCP/NFL}: \quad \left(g_{1*}, g_{2*}, g_{3*} \right) \simeq \left(9.4, -23.7, 0.5 \right) \times 10^{-3},
\end{align}
and annihilate for $N < \Nc$, leaving behind the runaway flow towards divergent short-range coupling. In contrast to the situation in $d=2+\epsilon$, the interaction channels now do not decouple, and all short-range couplings $g_\alpha$ therefore diverge at the same RG time. Their ratio, however, remains finite, and we find that $\lvert g_1/g_2 \rvert$ and $\lvert g_3/g_2 \rvert$ always remains small. Such hierarchy is usually taken as an indication that the dominant ordering tendency is the one that corresponds to the strongest short-range coupling~\cite{gehring2015}. In the present system, strong $g_2 < 0$ leads to nematic order~\cite{janssen2015a}. This suggests to associate the runaway flow with a nematic instability, as done in our previous work~\cite{herbut2014}.
The argument can be solidified by comparing susceptibilities in analogy to the analysis in the preceding section. We determine the flow of infinitesimally small ``mass parameters'' $\Delta_\alpha$ given in Eq.~\eqref{eq:L-Delta},
\begin{align}
 \frac{d \Delta_\text{nem}}{d \ln b} & = \left(z + \frac{2(e^2 + 2g_1)}{5} - \frac{4(4N + 3)g_2}{5}\right) \Delta_\text{nem}, \\
 \frac{d \Delta_\text{sc}}{d \ln b} & = \left(z - \frac{e^2+2g_1}{2} - 5g_2 \right) \Delta_\text{sc}, \\
 \frac{d \Delta_\text{ch}}{d \ln b} & = z \Delta_\text{ch}, \\
 \frac{d \Delta_\text{mag}}{d \ln b} & = \left(z + \frac{e^2+2g_1}{5} + \frac{2g_2}{5} \right) \Delta_\text{mag},
\end{align}
where we have neglected for simplicity terms $\propto g_3 \Delta_\alpha$ as those should give only small corrections of the order of $\lvert g_{3*}/g_{2*}\rvert \simeq 2\%$ to the flow of the $\Delta_\alpha$'s near the QCP, see Eqs.~\eqref{eq:g1-g2-g3-QCP} and \eqref{eq:g1-g2-g3-fixed-d}.
In the large-$N$ limit, we find the susceptibility exponents at the QCP as
\begin{align}
\gamma_\text{nem}/\nu & = 1 + \mathcal O(1/N), \\
\gamma_\text{sc}/\nu & = -1 + \mathcal O(1/N), \\
\gamma_\text{ch}/\nu & = -1 + \mathcal O(1/N), \\
\gamma_\text{mag}/\nu & = -1 + \mathcal O(1/N),
\end{align}
where $\nu = 1 + \mathcal O(1/N)$. The QCP therefore governs the continuous transition towards the nematic state, in agreement with the previous mean-field result~\cite{herbut2014}.
At the QCP-NFL merging point $(g_{1*}, g_{2*}, e^2_*)$ for $N \searrow \Nc$ the expansion is no longer under perturbative control, and the one-loop approximation does not necessarily lead to a unique positive susceptibility exponent. We find
\begin{align}
 \gamma_\text{nem}/\nu & \simeq -0.33, \\
 \gamma_\text{sc}/\nu & \simeq -1.32, \\
 \gamma_\text{ch}/\nu & \simeq -1.11, \\
 \gamma_\text{mag}/\nu & \simeq -0.95.
\end{align}
Although the one-loop result for $\gamma_\text{nem}/\nu$ is no longer positive, it still represents the largest value among the four examined here. This leads us to conclude that the runaway flow we find for $N < \Nc$ signals the onset of the nematic instability, in agreement with our results within the $2+\epsilon$ expansion.

\section{$4-\epsilon$ expansion} \label{sec:4-epsilon-expansion}

It has previously been shown that the properties of the Abrikosov-Beneslavskii NFL fixed point can be assessed by employing an $\epsilon$ expansion in $d=4-\epsilon$~\cite{abrikosov1974, moon2013, herbut2014}. Here, we demonstrate that the charged QCP that we found within the $2+\epsilon$ expansion (Sec.~\ref{sec:2+epsilon-expansion}) as well as the perturbative RG in fixed $d=3$ (Sec.~\ref{sec:perturbative-expansion}) can similarly be examined in a controlled way within a $4-\epsilon$ expansion.
To this end, we now focus on the nematic channel $\propto g_2$ in $L_\psi$ [Eq.~\eqref{eq:L-psi}] alone, which in both above approaches turned out to be the most dominant ordering tendency.

The quartic fermionic interaction can be traded for the corresponding Yukawa-type interaction by means of a Hubbard-Stratonovich transformation~\cite{janssen2015a},
\begin{equation}
 L_{\psi \phi} = h \phi_a \psi^\dagger_i \gamma_a \psi_i,
\end{equation}
where $\phi_a$, $a=1,\dots, (d/2+1)(d-1)$, represents the tensorial~\cite{notetensorial} nematic order-parameter field and $h$ is the Yukawa coupling.
RG loop corrections will generate a kinetic term for $\phi$ as well as bosonic self-interactions. We thus include these terms from the outset,
\begin{equation}
 L_{\phi} = \frac{1}{2} \phi_a \left(-c \partial_\tau^2 - \nabla^2 + r\right) \phi_a + \lambda \phi_a \phi_b \phi_c \Tr(\Lambda_{a}\Lambda_b\Lambda_c),
\end{equation}
where $\Lambda_a$ are the generalized real Gell-Mann matrices introduced in Eq.~\eqref{eq:hyperspherical}. The form of the cubic interaction parametrized by the coupling $\lambda$ is dictated by the rotational symmetry of the model~\cite{janssen2015a}. The flow of the parameter $c$ in front of the frequency term in $L_\phi$ is in general nontrivial and cannot be fixed to unity by simple rescaling. The boson mass $r$ can be understood as a tuning parameter for the nematic transition. The transition is signalled by a nonzero vacuum expectation value $\langle \phi_a \rangle \neq 0$, which is equivalent to $\langle \psi^\dagger_i \gamma_a \psi_i\rangle \neq 0$ for some $a$. The energetically favored direction of $\phi \equiv (\phi_a)$ leads to a uniaxial nematic state with a full gap in the fermionic spectrum~\cite{janssen2015a}. In $d=3$, it is given within our conventions by $\langle h \phi_5 \rangle > 0$ and $\langle \phi_1 \rangle, \dots ,\langle \phi_4 \rangle = 0$ [modulo $\mathrm{O}(3)$ rotations].
The resulting Gross-Neveu-Yukawa-type field theory is defined by the Lagrangian
\begin{equation} \label{eq:L-GNY}
 L = L_0 + L_a + L_{\psi \phi} + L_\phi.
\end{equation}
The theory is equivalent to the four-fermion model introduced in Eq.~\eqref{eq:L-2+eps} upon identifying
\begin{equation} \label{eq:g2-HST}
 g_2 \equiv -\frac{h^2}{2r}
\end{equation}
and setting $g_1 \equiv 0$ and $g_3 \equiv 0$ in the four-fermion theory, as well as setting $\lambda \equiv 0$ and taking the limit of $r\to \infty$ with fixed $h^2/r$ in the order-parameter theory.

The cubic coupling $\lambda$, the Yukawa coupling $h$, and the charge $e^2$ have engineering dimensions
\begin{equation}
 [h^2] = [\lambda^2] = [e^2] = 4-d.
\end{equation}
They thus become \emph{simultaneously} relevant below four dimensions. This suggests that the theory's critical behavior can be assessed within an $\epsilon$ expansion with small control parameter $\epsilon = 4-d$. Higher-order interactions are perturbatively irrelevant near $d = 4$ and have for this reason been omitted in $L$.
The analogous Gross-Neveu-Yukawa theory with the long-range Coulomb interaction neglected was investigated previously in Ref.~\cite{janssen2015a}. Here, we generalize these results to the case in which $e^2 > 0$.

Integrating the momentum shell from $\Lambda$ to $\Lambda/b$ leads to the flow equations
\begin{align}
 \frac{d e^2}{d \ln b} & = (z+2-d-\eta_a) e^2,
 \label{eq:beta-e2-GNY} \\
 \frac{d c}{d \ln b} & = (2-2z-\eta_\phi) c + \frac{2N}{5} h^2 + \frac{21}{4} \frac{ \sqrt{c}\lambda^2}{(1+r)^{5/2}},
 \label{eq:beta-c-GNY} \\
 \frac{d r}{d \ln b} & = (2-\eta_\phi)r - \frac{8N}{5} h^2 -
 21 \frac{\lambda^2}{\sqrt{c} (1+r)^{3/2}},
 \label{eq:beta-r-GNY} \\
 \frac{d h^2}{d \ln b} & = (6-d-z-\eta_\phi - 2\eta_\psi) h^2 + \frac{12}{5} \frac{h^4}{1+r} + \frac{4}{5} h^2 e^2,
 \label{eq:beta-h-GNY} \\
 \frac{d \lambda^2}{d \ln b} & = (6-d-z-3\eta_\phi) \lambda^2 - \frac{27}{2} \frac{\lambda^4}{\sqrt{c}(1+r)^{5/2}}
 \nonumber \\ & \quad
 - \frac{2\sqrt{3}N}{35} \lambda h^3,
 \label{eq:beta-lambda-GNY}
\end{align}
with the anomalous dimensions and the dynamical exponent as
\begin{align}
 \eta_a & = N e^2,
 \label{eq:eta-a-GNY} \displaybreak[0] \\
 \eta_\phi & = \frac{44}{35} N h^2 + \frac{21}{4} \frac{\lambda^2}{\sqrt{c}(1+r)^{5/2}},
 \label{eq:eta-phi-GNY} \displaybreak[0] \\
 \eta_\psi & = \frac{4}{5} \frac{h^2}{(1+r)^3} + \frac{4}{15} e^2,
 \label{eq:eta-psi-GNY} \displaybreak[0] \\
 z & = 2 - \eta_\psi.
 \label{eq:z-GNY}
\end{align}
Here, we have again employed the usual rescalings
\begin{gather}
 e^2 \Lambda^{d+\eta_a-z-2} S^d/(2\pi)^d \mapsto e^2, \\
 h^2 \Lambda^{d+z+\eta_\phi+2 \eta_\psi-6} S^d/(2\pi)^d \mapsto h^2, \\
 \lambda^2 \Lambda^{d+z+3\eta_\phi-6} S^d/(2\pi)^d \mapsto \lambda^2,
\end{gather}
and $c \Lambda^{2z+\eta_\phi-2} \mapsto c$, $\Lambda^{\eta_\phi-2} r \mapsto r$. Again, we have kept the general counting of dimensions in the couplings but have performed the angular integrations and the traces over spinor indices directly in $d=3$. $N$ consequently counts the number of four-component fermions $\psi_i$. Similarly, we have fixed the number of components of the nematic order-parameter field $\phi= (\phi_a)$ to be five in all dimensions, $a=1,\dots,5$. An alternative prescription to analytically continue the theory to noninteger dimension, in which the number of components of $(\phi_a)$ depends on $d$ leads to equivalent critical behavior, at least when $e^2 = 0$ and to the leading order in $\epsilon=4-d$~\cite{janssen2015a}.
We have also assumed $c$ to be small at the putative fixed point, $c_* = \mathcal O(\epsilon^\alpha)$ with $\alpha \geq 1$, which turns out to be consistent with the fixed-point values derived below.
Eqs.~\eqref{eq:beta-c-GNY}--\eqref{eq:eta-phi-GNY} and \eqref{eq:z-GNY} reduce to the flow equations listed in \cite{janssen2015a} when setting $e^2 \equiv 0$. Eqs.~\eqref{eq:beta-e2-GNY}, \eqref{eq:eta-psi-GNY}, \eqref{eq:eta-a-GNY}, and \eqref{eq:z-GNY} also agree with Ref.~\cite{moon2013} when setting $h \equiv 0$.

Similarly to the uncharged case~\cite{janssen2015a}, the QCP can readily be identified by introducing the new variables
\begin{align}
 u & = \frac{\lambda}{c_*^{1/4}}, &
 v & = \frac{h}{c_*^{1/12}},
\end{align}
with $c_*$ chosen such that it satisfies the fixed-point equation for $c$,
\begin{equation}
 0 = (2-2z)c_* + \left(\frac{2}{5} - \frac{44}{35}c_*\right) N c_*^{1/6} v^2.
\end{equation}
We therewith find an interacting charged fixed point to the leading order in $\epsilon = 4-d$ at
\begin{align}
 e^2_* & = \frac{15 \epsilon}{15N + 4} + \mathcal O(\epsilon^2), \displaybreak[0] \\
 c_* & = \mathcal O(\epsilon^{6/5}), \displaybreak[0] \\
 r_* & = \frac{6(5N+4)\epsilon}{15N+4} +  \mathcal O(\epsilon^{6/5}), \displaybreak[0] \\
 u_*^2 & = \frac{4(5N+4) \epsilon}{7(15N+4)} + \mathcal O(\epsilon^{6/5}), \displaybreak[0] \\
 v_*^2 & = \sqrt[3]{\frac{16(5N+4)(600N+515)^{2}}{21N}} \frac{\epsilon}{15N+4} + \mathcal O(\epsilon^{6/5}),
\end{align}
with $u_*$ and $v_*$ having opposite signs. Note that the two fixed points at $u_* > 0$, $v_* < 0$ and $u_*<0$, $v_* > 0$ are physically equivalent as the partition function is invariant under simultaneous sign change of both $h$ and $\lambda$.
As can be easily checked, the above fixed point is infrared attractive in the $(e^2, c, h, \lambda)$ coupling space. The only relevant direction is given by the tuning parameter $r$, and the fixed point hence indeed represents a QCP.
The corresponding critical exponents read
\begin{align}
 z & = 2 - \frac{4 \epsilon}{15N+4} + \mathcal O(\epsilon^{6/5}), \displaybreak[0] \\
 \eta_a & = \frac{15N \epsilon}{15N+4} + \mathcal O(\epsilon^{6/5}), \displaybreak[0] \\
 \eta_\psi & = \frac{4 \epsilon}{15N+4} + \mathcal O(\epsilon^{6/5}), \displaybreak[0] \\
 \eta_\phi & = \frac{(15N+12) \epsilon}{15N+4} + \mathcal O(\epsilon^{6/5}).
\end{align}
The correlation-length exponent is obtained from the flow of the tuning parameter
as
\begin{equation}
 1/\nu = 2 + \frac{15(5N+4) \epsilon}{15N+4} + \mathcal O(\epsilon^{6/5}).
\end{equation}
The nematic QCP that we have found within the $2+\epsilon$ expansion at large $N$ can hence be shown to exist also within the $4-\epsilon$ expansion by using a Gross-Neveu-Yukawa reformulation of the theory, which also allows to study the \emph{qualitative} properties of the nematic instability.
This represents the main result of this section.
We do not expect, however, that the \emph{quantitative} predictions when extrapolating our leading-order results to $\epsilon \to 1$ will accurately describe the physics in $d=3$. We therefore refrained from displaying the next-to-leading order corrections $\propto \epsilon^{6/5}$, which would be straightforwardly computable from the present one-loop flow equations.
The reason is that the QCP will interfere with the fully attractive NFL fixed point once we go sufficiently away from the upper critical dimension. This interplay is suppressed for small $\epsilon$, but not necessarily in $d=3$, as we have seen in Secs.~\ref{sec:2+epsilon-expansion} and \ref{sec:perturbative-expansion}, and as we will demonstrate within the Gross-Neveu-Yukawa formulation in Sec.~\ref{sec:frg}.

We now show that the NFL fixed point can also be rediscovered within the present formulation. This is achieved by employing a change of variables according to Eq.~\eqref{eq:g2-HST},
\begin{align} \label{eq:g-alpha-GNY}
 g & = - \frac{h^2}{2r},
 &
 \alpha & = \frac{r}{1+r},
\end{align}
with $g \leq 0$ and $0 \leq \alpha \leq 1$. (Here, we suppress the index of $g \equiv g_2$ for simplicity.) The upper bound $\alpha \nearrow 1$ corresponds to the limit of large boson mass $r \to \infty$, in which the order-parameter field decouples. $\alpha  \ll 1$ defines the quantum critical region with strong order-parameter fluctuations.
From Eqs.~\eqref{eq:beta-r-GNY} and \eqref{eq:beta-h-GNY} we find the flow equations in the new variables as
\begin{align}
 \label{eq:beta-alpha-GNY}
 \frac{d \alpha}{d \ln b} & = (1-\alpha) \left[
 (2-\eta_\phi)\alpha + \frac{16N}{5}g\alpha
 \right. \nonumber \\ & \quad \left.
 - 21(1-\alpha)^{5/2}\frac{\lambda^2}{\sqrt{c}}
 \right],
 \displaybreak[0] \\
 \label{eq:beta-g-GNY}
 \frac{d g}{d \ln b} & =
 (z-d) g
 - \frac{24}{5} \alpha g^2
 + \frac{4}{5} e^2 g
 - \frac{16N}{5} g^2
 \nonumber \\ & \quad
 + 21 \frac{\lambda^2}{\sqrt{c}}\frac{(1-\alpha)^{5/2}}{\alpha} g,
\end{align}
which exhibit a charged fixed point at $\alpha_* = 1$, $g_* = 0$, and $e^2_* = 15\epsilon/(15N+4)$. The fixed point is fully attractive in \emph{all} directions in coupling space, in particular, both $g-g_*$ and $\alpha-\alpha_*$ are irrelevant perturbations near the fixed point. The dynamical exponent is $z = 2 - 4\epsilon/(15N+4)$ and the Coulomb anomalous dimension is $\eta_{a} = 15N\epsilon/(15N+4)$, in precise agreement with the $4-\epsilon$ expansion results for the Abrikosov-Beneslavskii NFL fixed point given in Ref.~\cite{moon2013}.

We conclude that the present Gross-Neveu-Yukawa formulation of the theory allows to study the charged QCP and the NFL fixed point on an equal footing within an $\epsilon$ expansion around the upper critical spatial dimension of four. To the leading order in $\epsilon=4-d$, we find that both fixed points persist at real couplings for all $N \geq 1$.
However, as we shall see in the next section, once we go sufficiently away from the upper critical dimension, the two fixed points will approach each other upon lowering $N$ with $d$ held fixed (or equivalently upon lowering $d$ with $N$ held fixed).
Eventually, at some critical fermion number $\Nc(d)$ [equivalently, some critical dimension $d_\mathrm{c}(N)$] the fixed points will merge and annihilate. As a consequence, the flow from the weak-coupling regime with small $e^2>0$ and $h^2/(2r) \approx 0$, $\lambda \approx 0$ is ``bended'' from the NFL regime towards the symmetry-broken regime in which $r < 0$, leading to a nonvanishing vacuum expectation value $\langle \phi_a \rangle \neq 0$.

\section{FRG with dynamical bosonization} \label{sec:frg}

\begin{figure}[t]
 \includegraphics{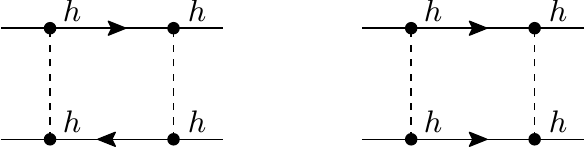}
 \caption{Feynman diagrams that generate four-fermion interactions from the Yukawa interaction at one-loop order. Solid (dashed) lines correspond to fermion (order-parameter) propagators.}
 \label{fig:box-phi}
\end{figure}

In order to arrive at the Gross-Neveu-Yukawa field theory in the previous section, we have traded the four-fermion term $g_2(\psi^\dagger_i \gamma_a \psi_i)$ for a Yukawa vertex $h \phi_a \psi^\dagger \gamma_a \psi$. While this allowed us to eliminate the four-fermion interaction at the ultraviolet scale, such terms will inevitably again be generated during the RG flow. At one loop, this happens by means of the box diagrams displayed in Figs.~\ref{fig:box-charge} and \ref{fig:box-phi}.
Within the $4-\epsilon$ expansion, these terms can be safely neglected as irrelevant, as done in the above. However, if we want to make contact with the results from $2+\epsilon$ expansion, these terms have to be taken into account, as they become marginal in $d=2$ and therefore potentially relevant at an interacting fixed point in $2<d<4$.
Fortunately, their influence can effectively incorporated into the present formulation by means of the so-called ``dynamical bosonization scheme''~\cite{gies2002}. The idea is to perform a Hubbard-Stratonovich transformation \emph{after every RG step}, such that newly generated four-fermion terms are always again converted into Yukawa interactions at each scale.
In this section, we implement this strategy entirely on the level of the standard Wilsonian momentum-shell RG. We understand it as a functional RG approach in the sense that perturbatively irrelevant operators are taken into account. In very much the same way, it can be implemented on the level of the flowing effective action~\cite{janssen2012}.

To be explicit, let us write down the effective action after integrating out a thin momentum shell between $\Lambda$ and $\Lambda/b$,
\begin{align} \label{eq:S-eff-FRG}
 S_{<} & = \int_{\vec k, \omega} \frac{1}{2}(r + \delta r)\phi_a^2
 + \int_{\vec k_1, \vec k_2, \omega_1,\omega_2} (h+\delta h)(\phi_a \psi^\dagger_i \gamma_a \psi_i)
 \nonumber \\ & \quad
 + \int_{\vec k_1, \vec k_2, \vec k_3, \omega_1,\omega_2,\omega_3}
 \delta g (\psi_i^\dagger \gamma_a \psi_i)^2
 + \dots,
\end{align}
where $\int_{\vec k, \omega} \equiv \int_0^{\Lambda/b}\! \frac{d \vec k}{(2\pi)^d} \int_{-\infty}^{\infty}\! \frac{d\omega}{2\pi}$, etc. The ellipsis represents the kinetic terms of $\psi$, $a$, and $\phi$, as well as the other interaction terms discussed in the previous sections, all of which do not play a direct role in the present discussion and as such are not explicitly displayed for notational simplicity. $\delta r=\mathcal O(\ln b)$ and $\delta h = \mathcal O(\ln b)$ denote the explicit loop corrections to the tuning parameter and the Yukawa vertex. They lead to the standard loop contributions to the flows of $r$ and $h$, as shown at one-loop order in Eqs.~\eqref{eq:beta-r-GNY} and \eqref{eq:beta-h-GNY}. $\delta g = \mathcal O(\ln b)$ describes the newly generated four-fermion term, which arises from the diagrams in Figs.~\ref{fig:box-charge} and \ref{fig:box-phi} and is assumed to be of the type corresponding to the proposed nematic instability.
The partition function reads
\begin{equation}
 \mathcal Z = \int\!\mathcal D \psi \mathcal D\psi^\dagger \mathcal D a \mathcal D \phi \,\ee^{-S_{<}(\psi,\psi^\dagger,a,\phi)}
\end{equation}
For any given configuration of $\psi$, $\psi^\dagger$, and $a$ we can shift $\phi_a$ in the inner functional integral $\int\!\mathcal D\phi \,\ee^{-S_<}$ as
\begin{equation}
 \phi_a \mapsto \phi_a + \delta \omega\,\psi^\dagger \gamma_a \psi,
\end{equation}
with arbitrary $\delta \omega = \mathcal O(\ln b)$. Then, for $\ln b \ll 1$ the partition function becomes
\begin{align} \label{eq:Z-dyn-bos}
 \mathcal Z & = \int\! \mathcal D \psi \mathcal D \psi^\dagger \mathcal D a \mathcal D\phi \,\exp \biggl\{ - \int_{\vec k,\omega,\dots} \biggl[ \frac{1}{2} (r+\delta r) \phi_a^2
 \nonumber \\ & \quad
 + (h+\delta h + r \delta \omega)(\phi_a \psi^\dagger_i\gamma_a\psi_i)
 \nonumber \\ & \quad
 + (\delta g + h \delta \omega)(\psi^\dagger_i \gamma_a \psi_i)^2
 + \dots
 \biggr] \biggr\}.
\end{align}
Thus, if we choose
\begin{equation}
 \delta \omega \equiv -\frac{\delta g}{h},
\end{equation}
the newly generated four-fermion terms will be exactly cancelled in Eq.~\eqref{eq:Z-dyn-bos}. At the same time, the Yukawa-coupling flow is modified as
\begin{equation}
 \frac{d h^2}{d \ln b} = (6-d-z-\eta_\phi-2\eta_\psi)h^2 + 2h \frac{\partial\,\delta h}{\partial \ln b} - 2r \frac{\partial\,\delta g}{\partial \ln b}.
\end{equation}
In the above equation, the second term represents the standard contribution from the explicit vertex renormalization, whereas the last contribution arises from the dynamical bosonization.
By computing the box diagrams in Figs.~\ref{fig:box-charge} and \ref{fig:box-phi} we find the modified flow
\begin{align} \label{eq:beta-h-FRG}
 \frac{d h^2}{d \ln b} & = (6-d-z-\eta_\phi - 2\eta_\psi)h^2 + \frac{12}{5} \frac{h^4}{1+r} + \frac{4}{5} h^2 e^2
 \nonumber \\ & \quad
 + \frac{13}{10}\frac{r h^4}{(1+r)^2} + \frac{r e^4}{10},
\end{align}
while the flow equations for $e^2$, $c$, $r$, and $\lambda$, as well as the anomalous dimensions $\eta_a$, $\eta_\phi$, $\eta_\psi$ and the dynamical exponent $z$ remain the same as in Eqs.~\eqref{eq:beta-e2-GNY}--\eqref{eq:z-GNY}.

\begin{table}[t]
 \caption{Fixed-point values and critical exponents at the QCP for different $N$ from functional RG.}
 \label{tab:exponents-QCP}
 \begin{tabular*}{\linewidth}{@{\extracolsep{\fill}}cccccccc}
 \hline\hline
 $N$ & $N e_*^2$ & $r_*$ & $N h_*^2$ & $\eta_{a}$ & $\eta_{\phi}$ & $2-z$ & $1/\nu$ \\ \hline
 1.856 & 0.87 & 12.75 & 1.45 & 0.87 & 1.82 & 0.13 & 0.00\\
 2 & 0.88 & 8.24 & 1.38 & 0.88 & 1.73 & 0.12 & 0.26 \\
 3 & 0.92 & 4.12 & 1.22 & 0.92 & 1.53 & 0.08 & 0.58 \\
 4 & 0.93 & 3.08 & 1.12 & 0.93 & 1.41 & 0.07 & 0.68 \\
 5 & 0.95 & 2.58 & 1.06 & 0.95 & 1.34 & 0.05 & 0.74 \\
 10 & 0.97 & 1.81 & 0.93 & 0.97 & 1.17 & 0.03 & 0.86 \\
 25 & 0.99 & 1.46 & 0.85 & 0.99 & 1.07 & 0.01 & 0.94 \\
 100& 1.00 & 1.32 & 0.81 & 1.00 & 1.02 & 0.00 & 0.99 \\
 $\infty$ & 1 & $\frac{14}{11}$ & $\frac{35}{44}$ & 1 & 1 & 0 & 1 \\ \hline \hline
 \end{tabular*}
\end{table}

Near the upper critical dimension of $d=4$, the additional two terms in Eq.~\eqref{eq:beta-h-FRG} as compared to Eq.~\eqref{eq:beta-h-GNY} are of subleading order both at the charged QCP as well as the NFL fixed point. Consequently, the fixed-point structure found in the previous section carries over completely to the present FRG approach when $d \nearrow 4$. Let us demonstrate that the dynamically bosonized flow coincides also with the flow as obtained within the $2+\epsilon$ expansion (Sec.~\ref{sec:2+epsilon-expansion}) in the limit $d \searrow 2$. To this end, we again introduce the variables $g$ and $\alpha$ as in Eq.~\eqref{eq:g-alpha-GNY}, leading to the modified flow equation
\begin{align}
  \frac{d g}{d \ln b} & =
 (z-d) g
 - \frac{24}{5} \alpha g^2
 + \frac{4}{5} e^2 g
 - \frac{16N}{5} g^2
 \nonumber \\ & \quad
 + 21 \frac{\lambda^2}{\sqrt{c}}\frac{(1-\alpha)^{5/2}}{\alpha} g
 -\frac{13}{5} \alpha^2 g^4 - \frac{e^4}{20},
\end{align}
and $(d \alpha)/(d \ln b)$ as in Eq.~\eqref{eq:beta-alpha-GNY}. For $\alpha \to 1$ we find that the above flow equation for $g$ agrees with Eq.~\eqref{eq:beta-g2} upon setting $g \equiv g_2$ and $g_1 = g_3 = 0$.
We therefore conclude that the approximation scheme is under perturbative control both near $d=2$ and near $d=4$.

\begin{table}[t]
 \caption{Fixed-point values and scaling exponents at the NFL fixed point for different $N$ from functional RG.}
 \label{tab:exponents-NFL}
 \begin{tabular*}{\linewidth}{@{\extracolsep{\fill}}cccccccc}
 \hline\hline
 $N$ & $N e_*^2$ & $r_*/N$ & $N h_*^2$ & $\eta_{a}$ & $\eta_{\phi}$ & $2-z$ & $\omega$ \\ \hline
 1.856& 0.87& 6.87 & 1.45 & 0.87& 1.82& 0.13 & 0.00 \\
 2& 0.88& 9.81 & 1.49 & 0.88& 1.88& 0.12 & 0.27 \\
 3& 0.92& 13.33 & 1.54 & 0.92& 1.94& 0.08 & 0.62 \\
 4& 0.94& 14.29 & 1.56 & 0.94& 1.96& 0.06 & 0.74 \\
 5& 0.95& 14.74 & 1.56 & 0.95& 1.97& 0.05 & 0.80 \\
 10& 0.97& 15.43& 1.58 & 0.97& 1.98& 0.03 & 0.91 \\
 25& 0.99& 15.74 & 1.59 & 0.99& 1.99& 0.01 & 0.96 \\
 100 & 1.00& 15.87 & 1.59 & 1.00 & 2.00 & 0.00 & 0.99 \\
 $\infty$ & 1 & $\frac{175}{11}$ & $\frac{35}{22}$ & 1 & 2 & 0 & 1 \\ \hline \hline
 \end{tabular*}
\end{table}
\begin{figure*}
\includegraphics[scale=1.04]{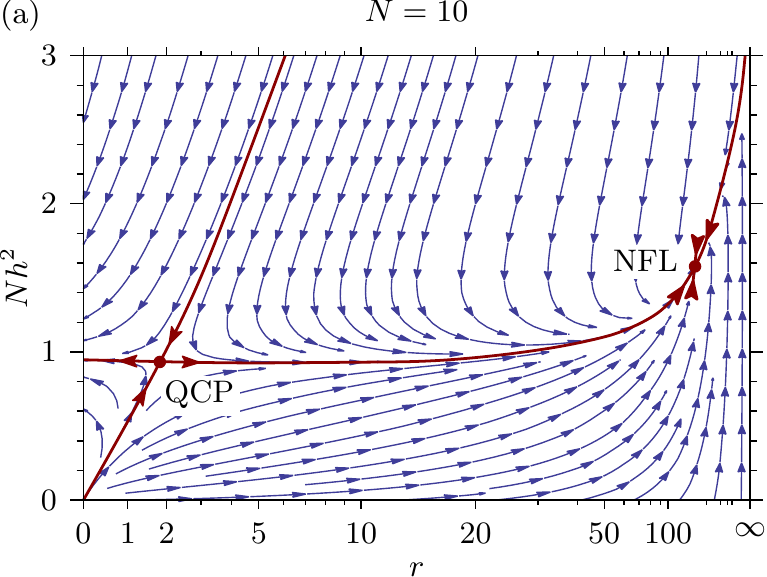}\hfill 
\includegraphics[scale=1.04]{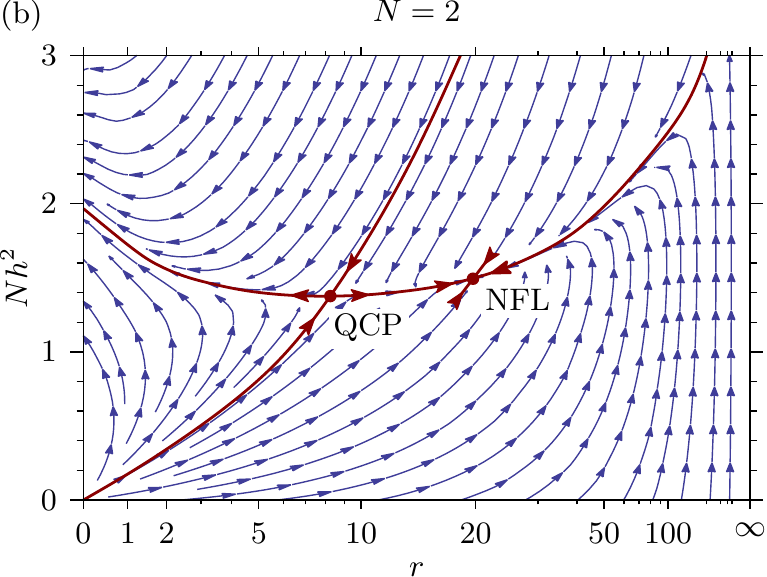}\\[\baselineskip]
\includegraphics[scale=1.04]{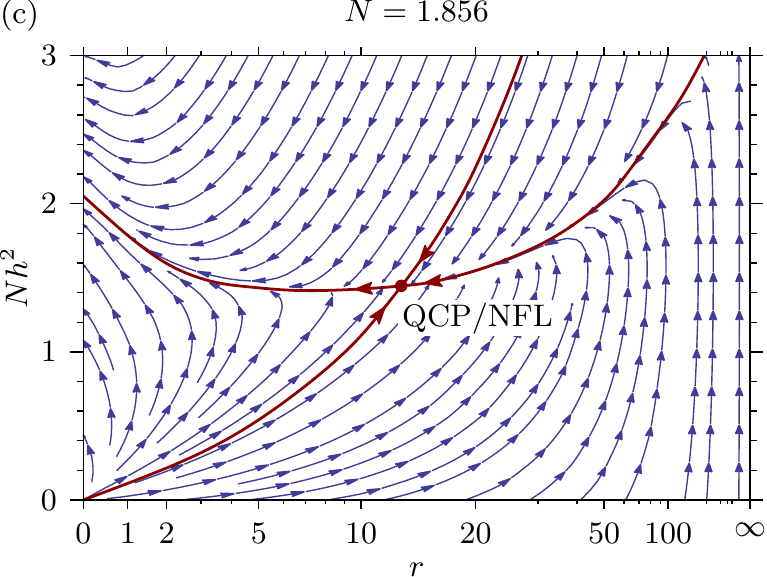}\hfill 
\includegraphics[scale=1.04]{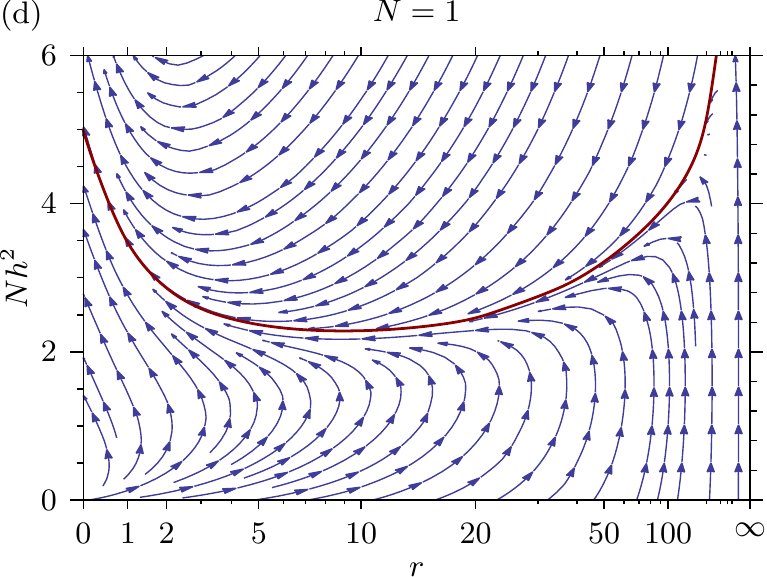}
\caption{RG flow diagram in $d=3$ for $N=10$, $2$, $1.856$, and $1$ in $r$-$h^2$ plane from functional renormalization group, displaying the quantum critical fixed point (QCP) and the fully attractive non-Fermi liquid fixed point (NFL). Arrows point towards infrared. To visualize the flow, the remaining couplings $c$, $e^2$, and $\lambda$ have been fixed at their values at the QCP. (For $N=1$, for which the QCP does not exist, we have chosen their large-$N$ predictions.) The horizontal axis has been rescaled by $r/(1+r)$ for reasons of clarity.
For $N>\Nc$ [(a),(b)] a weakly-correlated material with initially small coupling $\left. h^2/(2r) \right|_\text{UV} \ll 1$ flows to the fully attractive non-Fermi liquid (NFL) fixed point. A (hypothetical) strongly-correlated material with $\left. h^2/(2r) \right|_\text{UV} > \left(h^2/(2r)\right)_\mathrm{c}$ would flow to negative $r$, indicating a continuous~\cite{janssen2015a} phase transition towards a nematic state, with the critical behavior governed by the QCP. At $N \searrow \Nc = 1.856$ (c) the NFL fixed point merges with the QCP, such that for $N<\Nc$ (d) the flow is \emph{always} towards the nematic state, even for small initial coupling.}
\label{fig:FRG-flow}
\end{figure*}

We now turn to the physically interesting case of $d=3$.
In the limit of large $N$, the fixed-point equations can be solved analytically. In this limit, we recover both the QCP and the NFL fixed point. The former is located at
\begin{align}
\text{QCP}: \quad
\left(N e^2_*, c_*, r_*, N h_*^2, N \lambda_*^2\right) & =
\left(1, \tfrac{7}{22}, \tfrac{14}{11}, \tfrac{35}{44}, \tfrac{105}{85\,184}\right)
 \nonumber \\ & \quad
 + \mathcal O(1/N).
\end{align}
From this, we find $- h_*^2/(2r_*) = - 5/(16N) + \mathcal O(1/N^2)$, which agrees with the result of the fermionic formulation [Eq.~\eqref{eq:g1-g2-g3-QCP}] upon recalling the identification given in Eq.~\eqref{eq:g2-HST}.
The above fixed point has exactly one RG relevant direction, with the mass $r$ of the order-parameter field $\phi$ being the tuning parameter for the nematic transition.
The NFL fixed point is located in the large-$N$ limit at
\begin{align}
 \text{NFL}: \quad
 \left(N e^2_*, c_*, \tfrac{r_*}{N}, N h_*^2, N \lambda_*^2\right) & =
 \left(1, \tfrac{7}{44}, \tfrac{175}{11}, \tfrac{35}{22}, \tfrac{21}{13\,310}\right)
 \nonumber \\ & \quad
 + \mathcal O(1/N).
\end{align}
Note the different scaling of $r_*$ with $N$ as compared to the QCP: While the QCP is located at some finite $r_* = \mathcal O(1)$, the mass parameter at the NFL fixed point becomes large, $r_* = \mathcal O(N)$. Consequently, fluctuations of the order-parameter field are suppressed at the NFL fixed point for large $N$. (However, we shall see below that this is not necessarily the case for small $N$.)
At the NFL fixed point we have $-h_*^2/(2r_*) = - 1/(20N^2) + \mathcal O(1/N^2)$, which again precisely agrees with the large-$N$ result of the fermionic formulation, Eq.~\eqref{eq:g1-g2-g3-NFL}.
The NFL fixed point is RG attractive in all directions, with the corrections-to-scaling exponent $\omega$ that is determined by the RG flow in the ``least-irrelevant'' direction being $\omega = 1 + \mathcal O(1/N)$.

For finite $N$, we have solved the coupled system of fixed-point equations for $e^2$, $c$, $r$, $h$, and $\lambda$ numerically.
The results for the location of the fixed points and the corresponding universal scaling exponents are given for various $N$ in Tables~\ref{tab:exponents-QCP} and \ref{tab:exponents-NFL}. Note that the Coulomb anomalous dimension $\eta_a$ in each case fulfills the exact relation, Eq.~\eqref{eq:eta-a-exact}, as it should be.
With decreasing $N$, we find that the mass of the order-parameter field at the NFL fixed point rapidly decreases, thereby progressively enhancing fluctuations in the nematic channel. The QCP and NFL fixed point approach each other in coupling space, and eventually merge when $N \searrow \Nc$ with
\begin{equation}
 \Nc = 1.856.
\end{equation}
For $N < \Nc$ and small initial couplings $\left. h^2/(2r)\right\rvert_\text{UV} \approx 0$, $\left. \lambda\right\rvert_\text{UV} \approx 0$, and $0< \left. e^2 \right\rvert_\text{UV} \ll 1$, the RG flow is always towards the regime in which $r<0$, signalling the nematic transition and the spontanous breakdown of the rotational symmetry.
The RG flow for different values of $N$ above, at, and below $\Nc$ is visualized in Fig.~\ref{fig:FRG-flow}.

The fixed-point equations can be solved numerically for all $N$ in any given dimension $2<d<4$. We have displayed the result in terms of the critical fermion number $\Nc(d)$ in Fig.~\ref{fig:d-vs-Nfc}. As evident there, the FRG estimate approaches the result from the $2+\epsilon$ expansion for $d \searrow 2$, as expected. 
With the FRG, we can also estimate the critical dimension $d_\mathrm{c}$ below which the fixed-point annihilation occurs for fixed $N=1$. This way we find $d_\mathrm{c}(N=1) = 3.21$ and thus close to the value of $d_\mathrm{c} = 3.26$ found within the perturbative RG approach for $N=1$~\cite{herbut2014}.
For $d \nearrow 4$, we have explicitly checked that the scaling exponents for both the QCP and the NFL fixed point numerically coincide with their counterparts from the leading-order $4-\epsilon$ expansion. We reiterate that the FRG approach within the dynamical bosonization scheme becomes one-loop exact both in $d=2+\epsilon$ as well as in $d=4-\epsilon$, and smoothly interpolates between these two perturbatively accessible limits for intermediate dimension.

\section{Conclusions} \label{sec:conclusions}

In conclusion, we have studied the zero-temperature ground state of gapless semiconductors with quadratic Fermi nodes and weak long-range Coulomb interaction in 3D.
We have confined ourselves to a model in which the full rotational symmetry and the particle-hole symmetry is imposed from the outset, leaving the discussion of the subtle effects of these perturbations for a separate publication \cite{boettcher2}. The present  model serves as the minimal effective low-energy description of the electronic behavior of weakly-correlated 3D gapless semiconductors such as HgTe and $\alpha$-Sn~\cite{tsidilkovski1997}, and possibly also certain pyrochlore iridates of the form $R_2$Ir$_2$O$_7$ (with $R$ being a rare-earth element)~\cite{moon2013, kondo2015, nakayama2016}, as well as some half-Heusler compounds~\cite{chadov2010, lin2010, xiao2010}.
In order to gain analytical control over the low-temperature physics, we have extended the model by allowing a general number $N$ of fermion species (which may be understood as the number of QBTs at the Fermi level), as well as by generalizing to arbitrary spatial dimensionality $2<d<4$.
Our main results are the following:

\begin{enumerate}[(1)]
\item At large $N$, the system has a scale-invariant gapless ground state which is characterized by anomalous low-temperature power laws of various thermodynamic observables---a 3D non-Fermi liquid. The specific heat, for instance, would scale in this state as
\begin{equation} \label{eq:Cv-NFL}
 C_V \propto T^{d/z},
\end{equation}
with nontrivial dynamical exponent $1<z<2$. The emergence of the scale-invariant ground state can be traced back to the existence of a fully infrared stable NFL fixed point. We have computed the universal exponents in this state by employing $2+\epsilon$ expansion, perturbative RG in fixed $d=3$, $4-\epsilon$ expansion, and functional RG in the dynamical bosonization scheme. The results are consistent with the earlier works~\cite{abrikosov1974, moon2013}.

\item Upon lowering $N$, the NFL fixed point approaches another, quantum critical, fixed point. At some critical $\Nc$, the NFL fixed point and the QCP merge and eventually disappear for $N<\Nc$ into the complex-coupling plane.
As a consequence, the electronic system with a small number of QBTs at the Fermi level is unstable to weak long-range interactions.
Such fixed-point annihilation scenario was proposed earlier in the context of a $1/N$ expansion in fixed $d=3$~\cite{janssen2015b} as well as a one-loop RG for fixed $N=1$ and varying dimensionality~\cite{herbut2014}. In both these earlier approaches, however, the fixed-point annihilation occurs outside the regime in which the expansion is under control, and one may wonder whether higher loop orders may qualitatively change the conclusion. In the present work, we have demonstrated that the scenario can be described in a fully controlled way by employing an expansion around two spatial dimensions. On that point, we have exploited the fortunate fact that for small $\epsilon = d-2$ the fixed-point annihilation occurs at large $N$, thereby pushing the interesting physics entirely into the perturbative domain. This proves the existence of a phase boundary between the NFL phase at large $N$ and a novel symmetry-broken phase at small $N$ in the $d$-$N$ plane, see Fig.~\ref{fig:d-vs-Nfc}.

\begin{table}
 \caption{Critical fermion number $\Nc$ in $d=3$ spatial dimensions from different approaches.}
 \label{tab:Nc}
 \begin{tabular*}{\linewidth}{@{\extracolsep{\fill}}llr@{\extracolsep{0pt}}ll}
 \hline\hline
  Method & Reference & \multicolumn{2}{c}{$\Nc(d=3)$}\\ \hline
  $2+\epsilon$ expansion & Sec.~\ref{sec:2+epsilon-expansion} & 2.&56 \\
  RG in fixed $d=3$ & Sec.~\ref{sec:perturbative-expansion} & 2.&10 \\
  Functional RG & Sec.~\ref{sec:frg} & 1.&86 \\
  $1/N$ expansion in $d=3$ & Ref.~\cite{janssen2015b} & $\geq$ 2.&6(2) \\
  \hline\hline
 \end{tabular*}
\end{table}

\item Using a susceptibility analysis in $d=2+\epsilon$ dimensions, we find that the instability for $N<\Nc$ is towards a nematic state in which the rotational symmetry is spontaneously broken.
This result is in agreement with all other approaches employed here, as well as with the previous work~\cite{herbut2014}.
The low-$N$ quantum ground state has a full, but anisotropic gap and converts the semimetal into a topological Mott insulator.

\item We have employed a variety of different approaches to gain a reasonable estimate of the critical fermion number $\Nc$ in the limit of $d=3$. The results are summarized in Table~\ref{tab:Nc}. All estimates obtained so far consistently place the physical situation for $N=1$ into the Mott insulating regime. Gapless semiconductors with one quadratic Fermi node in 3D, such as clean HgTe and $\alpha$-Sn, should therefore suffer from a transition towards a nematic state in which an interaction-induced gap is dynamically generated. For such weakly-correlated materials the relevant energy scale at which interaction effects become important is $\varepsilon_* = \text{1--10\,meV}$~\cite{tsidilkovski1997, herbut2014}. From this and Eq.~\eqref{eq:gap} we estimate the size of the Mott gap $\Delta$ for $\Nc \approx 2$ and $z \approx 1.7$ as
\begin{equation}
 \Delta \approx 10^{-2}\varepsilon_* \approx \text{0.01--0.1\,meV}.
\end{equation}
As function of temperature, we correspondingly expect the Mott transition to occur at a critical temperature $T_\mathrm{c}$ of the order $T_\mathrm{c} \approx \text{0.1--1\,K}$, in agreement with~\cite{herbut2014}. Besides the opening of the Mott gap, the transition reveals itself experimentally through a thermodynamic singularity at $T_\mathrm{c}$, as measurable, for instance, in the specific heat. Below the jump at $T_\mathrm{c}$, the specific heat is exponentially suppressed,
\begin{equation}
 C_V \propto \ee^{-\Delta/(2k_\mathrm B T)},
\end{equation}
while for $T > T_\mathrm{c}$ it is expected to resemble the NFL behavior with nontrivial exponents as in Eq.~\eqref{eq:Cv-NFL}, see Ref.~\cite{herbut2014}. The Hall coefficient should exhibit a similar temperature dependence~\cite{janssen2015b}. The effects are experimentally accessible if the sample can be prepared sufficiently pure.

\end{enumerate}

To approach the complex physics in the pyrochlore iridates, the nontrivial interplay of the (itinerant) iridium electrons with the (local) rare-earth magnetic moments should be investigated~\cite{nakayama2016}.  A separate paper \cite{boettcher2} will present a detailed discussion of the effects of deviations from the spherical and particle-hole symmetries which may also be important for this class of materials \cite{savary2014, goswami2016}.

\begin{acknowledgments}
We thank I.~Boettcher, H.~Gies, and T.~Senthil for discussions. This work was supported by the DFG through JA2306/1-1, JA2306/3-1, and SFB 1143, as well as the NSERC of Canada.
\end{acknowledgments}

\end{document}